\documentclass[english]{article}
\usepackage[T1]{fontenc}
\usepackage{amsmath}
\usepackage[latin9]{inputenc}
\usepackage{geometry}
\usepackage{color}
\definecolor{note_fontcolor}{rgb}{0.800781, 0.800781, 0.800781}
\usepackage{amstext}
\usepackage{graphicx}
\usepackage{esint}
\usepackage{color}
\usepackage{authblk}
\usepackage{amsthm}
\usepackage{cite}
\usepackage{url}
\usepackage{float}
\usepackage{amsmath,amssymb,amsthm}
\usepackage{longtable}

\usepackage{lineno,xcolor}

\definecolor{dkgreen}{rgb}{0,0.6,0}

\makeatletter

\newcommand{\lyxmathsym}[1]{\ifmmode\begingroup\def\b@ld{bold}
  \text{\ifx\math@version\b@ld\bfseries\fi#1}\endgroup\else#1\fi}

\usepackage[font={small,it}]{caption}

\makeatother

\usepackage{babel}

\newtheorem{theorem}{Theorem}
\newtheorem{mydef}{Definition}

\begin{document}

\title{Effect of antipsychotics on community structure in functional brain networks}
\author{Ryan Flanagan$^1$, Lucas Lacasa$^1$, Emma K. Towlson$^2$, Sang Hoon Lee$^{3}$, and Mason A. Porter$^{4}$}
\affil{\small{$^1$School of Mathematical Sciences, Queen Mary University of London, London E1 4NS, UK\\
$^2$Center for Complex Network Research and Department of Physics, Northeastern University, Boston, MA 02115, USA\\
$^3$Department of Liberal Arts, Gyeongnam National University of Science and Technology, Jinju 52725, Korea\\
$^4$Department of Mathematics, University of California, Los Angeles, CA 90095, USA}} 


\maketitle


\begin{abstract} 

Schizophrenia, a mental disorder that is characterized by abnormal social behavior and failure to distinguish one's own thoughts and ideas from reality, has been associated with structural abnormalities in the architecture of functional brain networks. Using various methods from network analysis, we examine the effect of two classical therapeutic antipsychotics --- Aripiprazole and Sulpiride --- on the structure of functional brain networks of healthy controls and patients who have been diagnosed with schizophrenia. We compare the community structures of functional brain networks of different individuals using mesoscopic response functions, which measure how community structure changes across different scales of a network. We are able to do a reasonably good job of distinguishing patients from controls, and we are most successful at this task on people who have been treated with Aripiprazole. We demonstrate that this increased separation between patients and controls is related only to a change in the control group, as the functional brain networks of the patient group appear to be predominantly unaffected by this drug. This suggests that Aripiprazole has a significant and measurable effect on community structure in healthy individuals but not in individuals who are diagnosed with schizophrenia. In contrast, we find for individuals are given the drug Sulpiride that it is more difficult to separate the networks of patients from those of controls. Overall, we observe differences in the effects of the drugs (and a placebo) on community structure in patients and controls and also that this effect differs across groups. We thereby demonstrate that different types of antipsychotic drugs selectively affect mesoscale structures of brain networks, providing support that mesoscale structures such as communities are meaningful functional units in the brain.

\end{abstract}



\section{Introduction\label{sec:Intro}}

Investigating the structure and dynamics of neuronal networks is crucial for understanding the human brain, and the nascent field of ``network neuroscience'' has yielded fascinating insights into a diverse variety of neurological phenomena~\cite{bassett2017,Betzel2016}. Recent advances in imaging technology have made it possible to perform increasingly detailed analyses of brains, and it is now possible to map anatomical regions and their interconnecting pathways at near-millimeter resolution. This yields large-scale networks with which to describe the brain's structural connectivity (i.e., the human connectome)~\cite{sporns2005human,sporns2013structure}. These structural connections govern large-scale neuronal dynamics, which can be captured as patterns of functional connectivity in so-called ``functional brain networks''~\cite{sporns2015cerebral,sporns2015modular,papo2014complex,Petersen2015207}. Such functional networks are usually built by estimating dynamical correlations or other interdependencies in the neuronal activity of brain regions. 

One can measure functional brain networks using various approaches, such as through blood oxygen level dependent (BOLD) signals gathered via functional magnetic resonance image (fMRI) scans or using other modalities~\cite{bassett2017,Betzel2016,papo2014complex,sporns2013structure}.
Such studies have yielded many fascinating insights, such as successful detection of irregularities in fMRI data of patients with disorders and diseases like Alzheimer's disease~\cite{supekar2008network}, autism~\cite{dichter2012functional}, schizophrenia~\cite{lynall2010functional,van2013abnormal,hadley2016change}, and others~\cite{greicius2008resting}. 

In the present paper, we seek to analyze the specific role and effect that different antipsychotics (specifically, Aripiprazole and Sulpiride) play in the structure --- especially community structure, in which densely-connected sets of nodes are relatively sparsely connected to other densely-connected sets of nodes~\cite{Porter2009,Fortunato2016,Fortunato2010} --- of functional brain networks of both control and patients. Schizophrenia is often characterized by abnormal and inconsistent social behavior and failure to differentiate between thoughts and reality. Methods for diagnosing schizophrenia have been somewhat controversial~\cite{kendell2014distinguishing}, and scientists and doctors seek to understand and develop effective diagnoses and treatment (in the form of therapy and drugs)~\cite{kay1987positive}. The atypical antipshchotic drug {\it Aripiprazole}, which acts as a partial dopamine agonist, is used primarily for the treatment of schizophrenia. {\it Sulpiride}, another atypical antipsychotic that works as a selective dopamine agonist, is also used to treat schizophrenia. Their effectiveness has been reported widely in the literature, and their use for treatment has been approved in many countries~\cite{marder2003aripiprazole,shiloh1997sulpiride}. (We note, however, that Sulpiride is not approved for use in the United States, Canada, or Australia.)

It has been hypothesized that schizophrenia is related to abnormalities in the connectivity between components of functional brain networks~\cite{lynall2010functional}. Furthermore, although the biological mechanisms of Aripiprazole and Sulpiride are well-understood, their effects at the functional level of the brain are not. This motivates our goal to explore the fingerprint of such drugs in functional brain networks. An important property of a functional brain network, which appears to be abnormally altered in patients diagnosed with schizophrenia, is community structure~\cite{alexander2012discovery,Betzel2016}. Loosely speaking, a community is a set of nodes in a network that are connected densely to each other but connected sparsely to other parts of a network~\cite{Porter2009,Fortunato2016}. Community structure in a network is one type of mesoscale organization, and both community structure and other mesoscale organizations (e.g., core--periphery structure~\cite{bassett2013task}) are important in a variety of contexts in functional brain networks~\cite{Betzel2016}.

To examine community structure in healthy individuals versus individuals with schizophrenia under the effects of different drugs, we employ several characterizations of graph similarity. We consider both basic features (such as the number of common edges) and more sophisticated ones (such as how community structure changes over different scales of a network~\cite{onnela2012taxonomies}). This suite of techniques allows us to build a set of distance matrices between subjects, and we apply unsupervised clustering algorithms to these matrices to try to identify discernible groups of subjects. We focus in particular on studying the effects of each drug within a given group (intra-subject comparisons), though we also compare groups under the same drug (inter-subject comparisons). We thereby investigate both the difference between healthy subjects and patients and the effects that each of the drugs have on the functional brain networks of each group of subjects.

The rest of our paper is organized as follows. In Section~\ref{sec:Data}, we briefly discuss the data set and some relevant previous studies. In Section~\ref{sec:Methods}, we detail the protocol and the methods that we use to make comparisons between groups of subjects. In Section~\ref{sec:Results}, we present our results. Finally, in Section~\ref{sec:Discussion}, we discuss the implications of our findings. In an appendix, we state and prove a theorem (that a certain diagnostic has a metric structure) that we use in the main text.


\section{Data and Previous Studies} \label{sec:Data} 

The data set, which came from Bristol Myers Squibb (BMS) and which we call the ``BMS data set'', consists of measurements of 15 healthy human subjects (``controls'') and 12 human subjects (``patients'') who were diagnosed previously with schizophrenia. All participants were pre-treated with domperidone on all three days to reduce side effects. Over 3 sessions, which were 1--2 weeks apart, each of the 27 subjects was given one of three different drug treatments:
\begin{enumerate}
\item (``Placebo'') Oral placebo, 180 and 90 minutes before scanning;
\item (``Sulpiride'') Oral placebo, 180 minutes before scanning; and then oral Sulpiride (400 mg), 90 minutes before scanning;
\item (``Aripiprazole'') Oral Aripiprazole (15 mg), 180 minutes before scanning; and then oral placebo, 90 minutes before scanning.
\end{enumerate}
At each session, after being given one of the drug treatments, each individual was placed in an fMRI scanner to measure blood flow in the brain. The fMRI scanner captures a single image once every 2 seconds. The scans lasted 17 minutes and 4 seconds, so each BOLD time series has 512 time points. The data are parcellated into 298 regions of interest (RoIs), and each region corresponds to a node in a functional brain network. The parcellation is based on an existing method, but with 27 regions removed due to problems with head motion~\cite{zalesky2010whole, patel2014wavelet}. Each region has a corresponding time series that represents an average level of activity in that region. We remove 4 controls (2, 8, 10, and 14) and 3 patients (3, 5, and 11) from our calculations due to missing data and/or problems due to head motion. We thus examine a total of 20 subjects: 11 controls and 9 patients. See \cite{power2012,satt2012,vandijk2012} for discussions of issues with head motion, and see \cite{suckling2006,patel2014wavelet} for discussions of preprocessing of fMRI data to correct for head motion.

There have been three previous studies~\cite{lynall2010functional,zalesky2012relationship,EmmaPaper} that employed this particular data set. They dealt solely with the task of distinguishing controls from patients who had been diagnosed with schizophrenia, so they were trying to find effective biomarkers for schizophrenia. Using a parcellation with 90 RoIs, Ref.~\cite{lynall2010functional} reported that the patients have ``less strongly connected'' brain networks (i.e., in the sense of a lower mean pairwise wavelet coherence between regions) and ``more diverse'' profiles (in the sense of larger mean variances in a wavelet coherence between a given region and the others) of brain functional connectivity than the controls. They also calculated that brain networks in the schizophrenia group have a greater robustness to uniform-at-random removal of nodes, in the sense that the number of nodes in the largest connected component (LCC) decays more slowly as a function of the number of removed nodes. Reference~\cite{zalesky2012relationship} built functional networks via ``spatial pairwise clustering'' (a novel approach that they introduced) of individual voxels (thereby foregoing the need to choose a parcellation) and combining spatially proximate voxels into nodes. In their computations, they observed weaker inter-nodal correlations in patients than in controls. Finally, using a very similar parcellation to the one that we employ, a very recent work \cite{EmmaPaper} studied the effects of the drugs on the networks of the subjects. Their results suggest that (1) Aripiprazole has a major effect on the networks of healthy people and that (2) both drugs make it harder to distinguish controls and patients.


\begin{figure}
\begin{centering}
\includegraphics[scale=0.35]{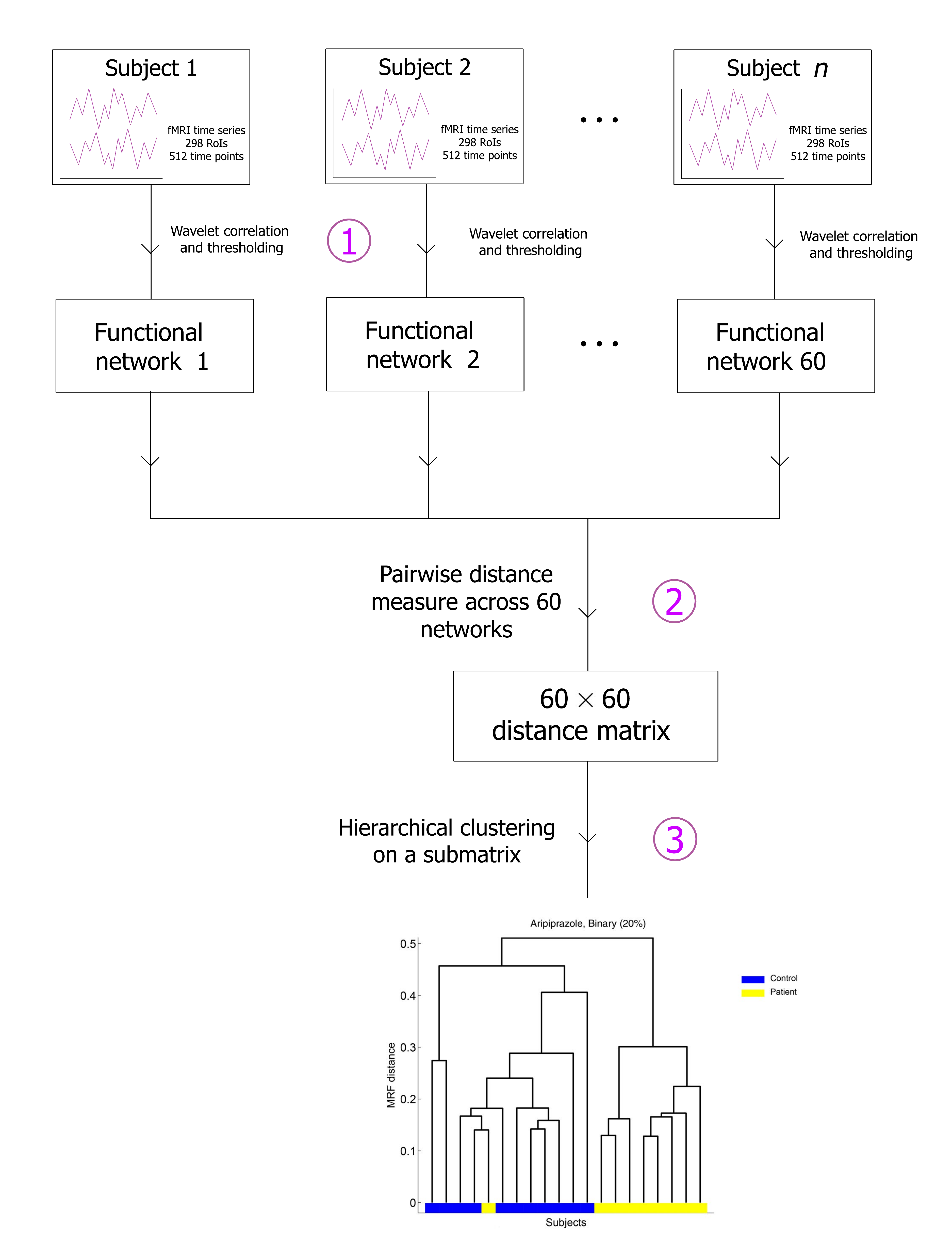}
\par\end{centering}
\protect\caption{\label{fig:PaperProcess} Protocol to obtain a dendrogram that conveys hierarchical clustering of a set of subjects. The submatrix that we use depends on our particular comparison from Fig.~\ref{fig:Comparisons-1}. We explain the vertical axis in the dendrogram in Section \ref{CommunityDistance}. In the example dendrogram in this schematic, we consider unweighted networks that the include strongest 20\% edges (see Section \ref{sub:CreateNetworks}).
}
\end{figure}


\section{Methods and Preliminary Computations} \label{sec:Methods}

We illustrate our analysis pipeline in Fig.~\ref{fig:PaperProcess}. In Subsections \ref{sub:CreateNetworks} and \ref{scale}, we briefly describe how to build a functional network from fMRI time series using wavelet correlations and thresholding techniques (step 1 in Fig.~\ref{fig:PaperProcess}). In Subsection \ref{sub:Simple-Network-Measures}, we discuss our preliminary computations on the collection of networks. In Subsections \ref{sub:Similarity-and-Distance} and \ref{sub:Clustering}, we discuss how to define two distance functions to examine dissimilarities of functional networks (step 2 in Fig.~\ref{fig:PaperProcess}) and how to apply hierarchical clustering to cluster similar subjects (i.e., similar functional networks) according to step 3 in Fig.~\ref{fig:PaperProcess}).


\subsection{Building the Networks} \label{sub:CreateNetworks}

Wavelet-based correlations allow one to examine functional similarities between brain regions based on activity in a specified frequency interval (a so-called wavelet ``scale''). We use the maximal-overlap discrete wavelet transform~\cite{percival2000wavelet} to decompose each regional mean fMRI time series (see step 1 in Fig.~\ref{fig:PaperProcess}). Examining wavelets is useful for studying resting-state fMRI data, and functional connectivity between regions is typically largest at certain frequency bands (below 0.1 Hz)~\cite{cordes2000mapping}. Let $g_{i}$ denote the time series of node (i.e., RoI) $i$ (where $i \in \{1,2,\dots, 298\}$), and let $V_{s}(g_{i})$ denote the vector of scale-$s$ wavelet coefficients of $g_{i}$. At scale $s$, the connection strength between two nodes $i$ and $j$ in a functional network is given by the wavelet correlation
\begin{equation}
	F_{ij}=\frac{\sum_{k}V_{s,k}(g_{i})V_{s,k}(g_{j})}{\sqrt{(\sum_{k}(V_{s,k}(g_{i}))^{2}(\sum_{k}(V_{s,k}(g_{j}))^{2}}} \in [-1,1]\,.
\label{eq:F_ij}
\end{equation} 
Fe compute values of $F_{ij}$ for scales $s=1,\,2,\,3,\,4$; and we then choose to work with the most informative scale (see Section~\ref{sec:Results}).

There are $N=298$ RoIs for each subject, so we extract functional networks with $N=298$ nodes. This yields a similarity matrix $\mathbf{F}$ whose elements are given by Eq.~\eqref{eq:F_ij}. To avoid negative weights\footnote{There are also other ways to transform $\mathbf{F}$ into a weighted adjacency matrix ${\bf W}$. For example, one can take the absolute value of the correlations, though it is then impossible to distinguish negative wavelet similarities from positive ones. The weakness of our chosen approach is that initially strongly negative weights are transformed into weights that are near $0$, and they then tend to be removed if one subsequently prunes a network by keeping only the most strongly weighted edges of ${\bf W}$. Recently, \cite{zhan2017significance} examined the significance of such negative wavelet similarities.}, we transform $\mathbf{F}$ into a weighted adjacency matrix $\bf{W}$ by taking $W_{ij}:=(F_{ij}+1)/2 \in [0,1]$. The associated network is fully connected by construction, and there are two customary ways to prune edges. These are (1) thresholding the networks by keeping a fixed fraction $\tau$ of the strongest weights (assigning the remaining edges a weight of $0$ and producing thresholded weighted networks) and (2) first performing the previous step and then subsequently setting the remaining edges a weight of $1$, thereby producing thresholded binary networks. In both cases, the resulting thresholded networks have $E=N(N-1)\tau/2$ edges. Of course, one can also simply keep all edges and examine the original fully connected, weighted networks.
In the present paper, we initially examine the original networks and both the weighted and binary thresholded networks. Based on some preliminary calculations, we then decide which of these networks to examine further. 


\begin{figure}
\begin{centering}
\includegraphics[scale=0.9]{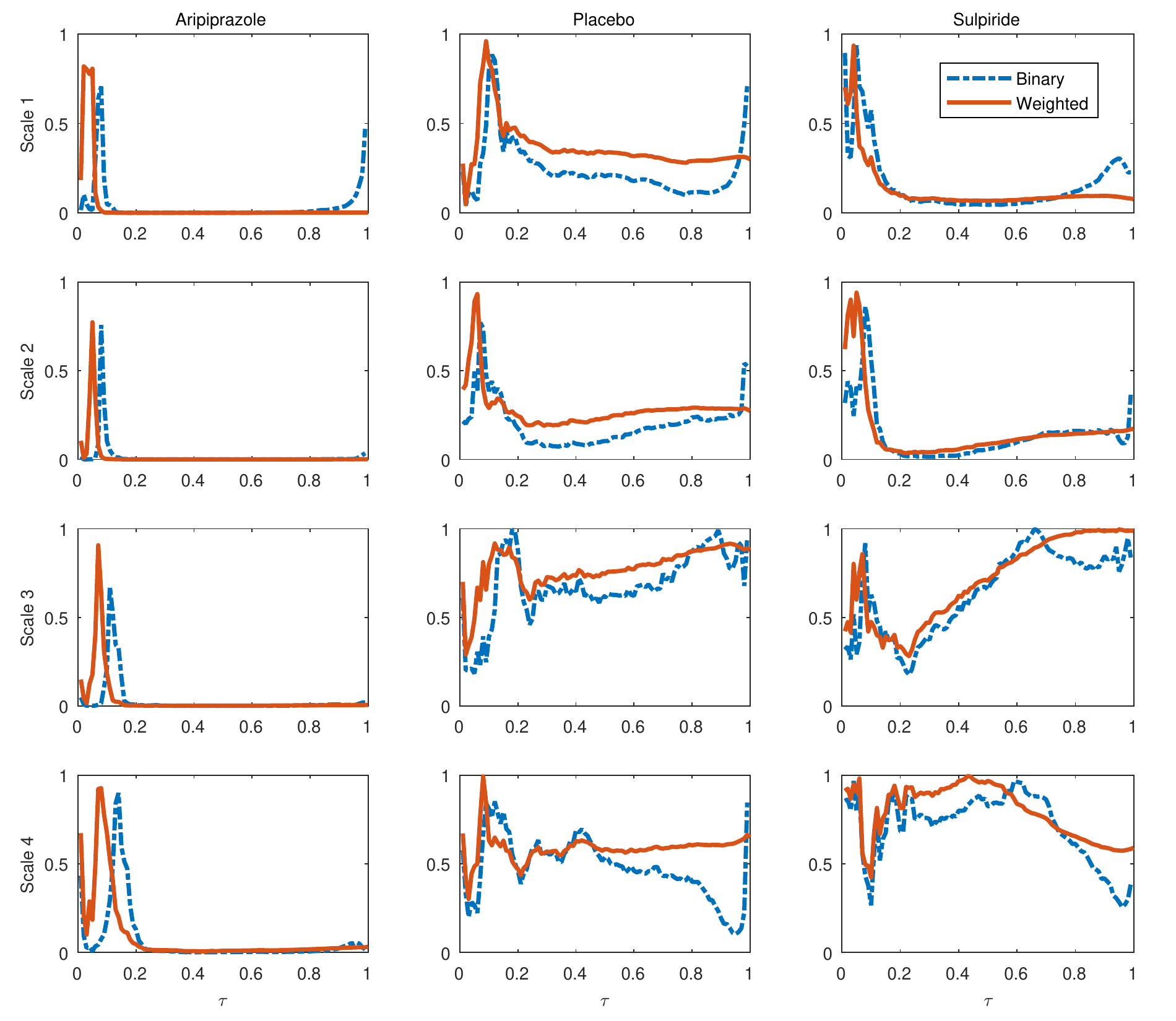}
\par\end{centering}
\protect\caption{\label{fig:Clustering}The p-values associated with $t$-tests on the mean local clustering coefficient (between healthy patients and controls) for weighted networks (solid red curves) and binary networks (blue dashed curves) for different values of the thresholding parameter $\tau$. Wavelet scale 2 produces the smallest p-values. We also observe a difference in the curves of the three drug treatments and that the p-values associated with the binary networks are consistently smaller than those for the weighted ones.
}
\end{figure}


\subsection{Choosing a Scale and Thresholding Parameter} \label{scale}

To construct the functional networks, we choose a scale $s$ and then consider thresholding the networks (with an associated threshold value). Previous work has noted differences in both ``connectivity'' (i.e., the mean edge weight of a network) and mean local clustering coefficient between healthy controls and patients with schizophrenia \cite{lynall2010functional,liu2008disrupted,he2012altered}. The observed difference were more statistically significant at lower frequencies, and it was particular evident at scale 2. 
This is consistent with previous research on resting-state fMRI \cite{achard2006resilient}. To make an educated choice of scale, Ref.~\cite{lynall2010functional} calculated the mean value of $F_{ij}$ over healthy controls and patients for each scale, performed a $t$-test, and selected the scale with the smallest p-value. We follow a similar procedure, but we also threshold the networks for both binary and weighted versions using a thresholding parameter $\tau$, in which we keep a fraction $\tau$ of the strongest edges.\footnote{We consider values of $\tau$ in increments of $0.01$.}
 (For example, if $\tau=0.4$, we keep the strongest 40\% of the edges.) For each of the three drug treatments and for each of the scales 1, 2, 3, and 4, we then perform a $t$-test on the mean local clustering coefficients of healthy controls and patients diagnosed with schizophrenia. In Fig.~\ref{fig:Clustering}, we show all 12 plots and the p-values associated with the $t$-tests. Based on these results, we make two decisions. First, from now on, we use scale 2 (which corresponds to the frequency band 0.060--0.125 Hz), because it has the smallest p-values (in agreement with previous work \cite{lynall2010functional}). For very small values of $\tau$, we observe spikes in the p-values that likely arise from the networks breaking up into many components. Second, because our results on binary networks have smaller p-values than the corresponding ones for weighted networks, we focus our subsequent calculations on thresholded binary networks (except for our calculations of connectivity). The controls tend to have much larger edge weights than the patients, so our comparisons between patients and controls are more directly parallel if we use binary networks, as many network quantities are affected in nontrivial ways by edge weights. From now on, we fix $\tau = 0.2$. (We repeat our calculations for values of $\tau \in [0.2,0.4]$, and we obtain qualitatively similar results.) 

Half of our networks (30 out of 60) have more than one component when $\tau=0.2$. This can be problematic for some types of computations, such as those that involve path lengths, but this issue has not posed a problem in practice in the present work, and the largest connected component of every network has at least 291 nodes (out of 298 nodes in total). In Appendix \ref{appendix:components}, we show the number and sizes (i.e., number of nodes) of the components in each of our networks.


\subsection{Connectivity and Mean Local Clustering Coefficient} \label{sub:Simple-Network-Measures}

We now do some preliminary calculations. Previous research using thresholded, binary networks has highlighted significant differences in ``connectivity'' (defined, for an individual subject, as the mean edge weight $\langle W_{ij} \rangle$ of a network) and mean local clustering coefficients of networks from control subjects versus those from patients diagnosed with schizophrenia~\cite{lynall2010functional}. In our case, by construction, connectivity corresponds (up to a scaling and a shift) to the mean wavelet correlation. For weighted networks, we compute the weighted local clustering coefficient\cite{saramaki2007generalizations} 
\begin{equation} \label{cc}
	c_{i}=\frac{1}{k_{i}(k_{i}-1)}\sum_{j,k}(W_{ij}W_{ik}W_{jk})^{1/3} \quad \textrm{for } k_i \geq 2\,,
\end{equation}
where $k_i$ is the degree of node $i$, and $c_i = 0$ if $k_i \in \{0,1\}$. Equation~\eqref{cc} reduces to the usual local clustering coefficient for the special case of binary networks.

\begin{figure}[H]
\begin{minipage}[t]{0.45\textwidth}%
\begin{center}
\includegraphics[scale=0.41]{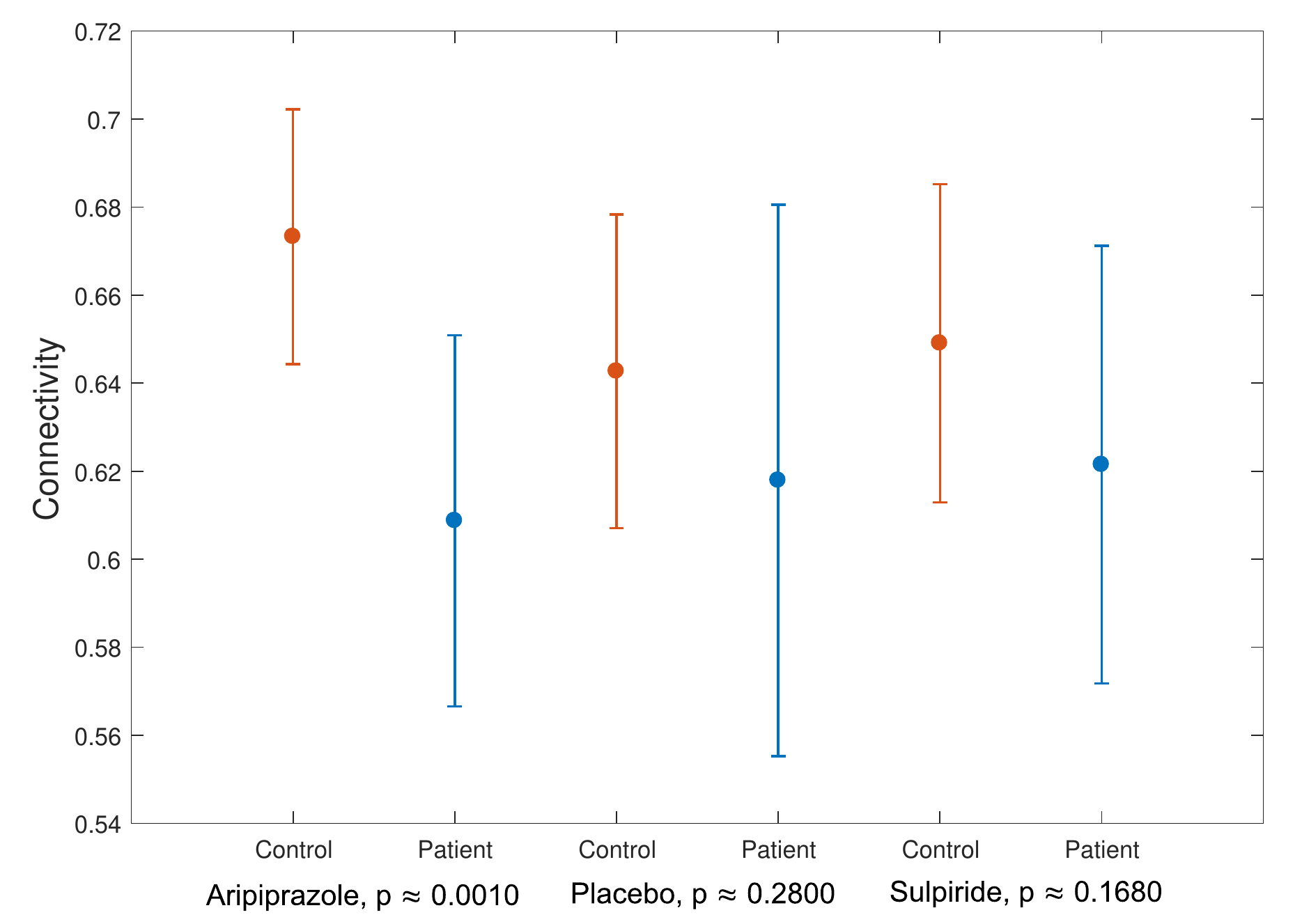}
\par\end{center}%
\end{minipage}\hfill{}%
\begin{minipage}[t]{0.45\textwidth}%
\begin{center}
\includegraphics[scale=0.41]{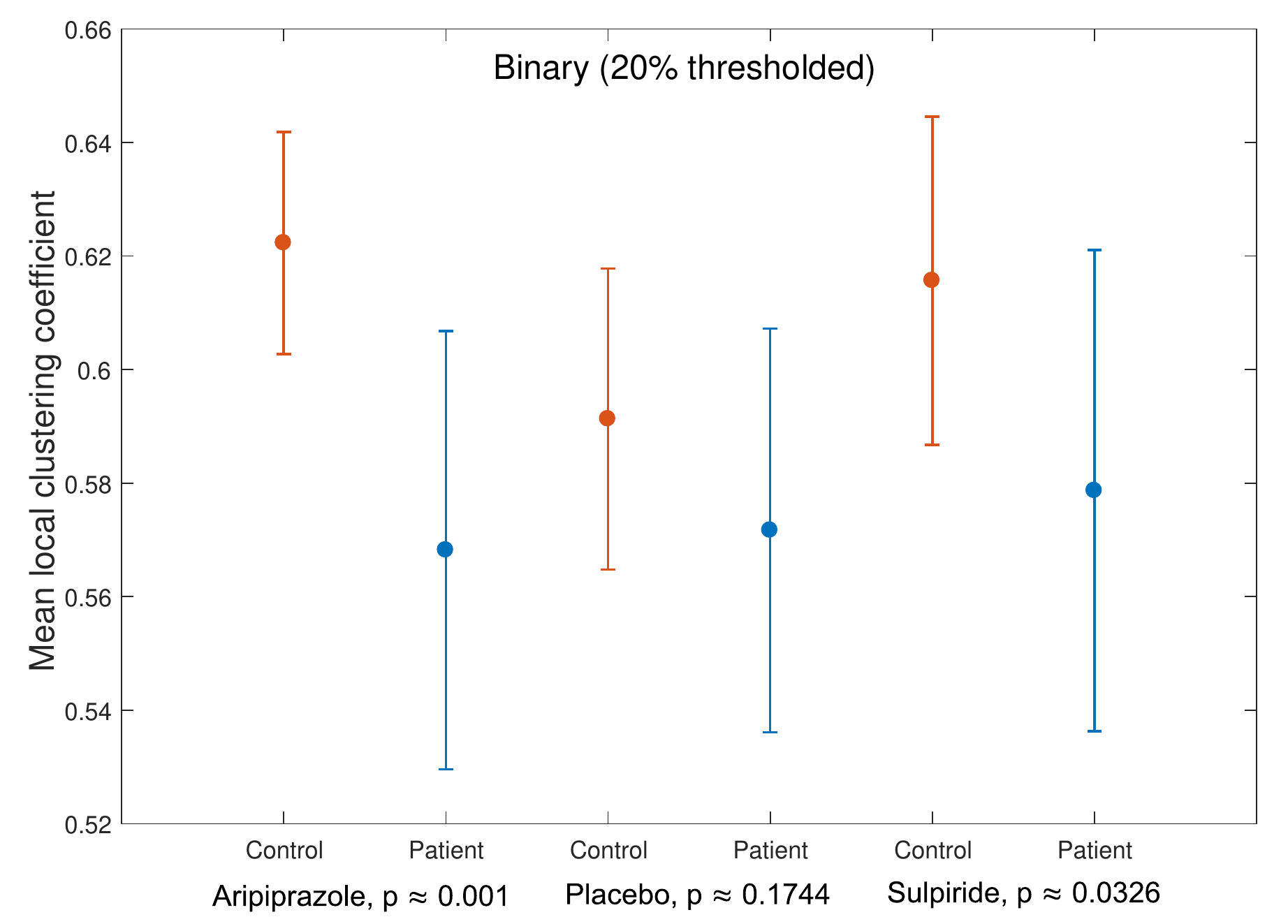}
\par\end{center}%
\end{minipage}\protect\caption{\label{fig:clustering}Means and standard deviations of (left) connectivity for non-thresholded weighted networks and (right) mean local clustering coefficients for binary networks thresholded to $20\%$ of the strongest edges. The results are similar in each case, though for Sulpiride we observe a difference between controls and patients in the p-value for the two-sample \textit{t}-test.
}
\end{figure}

\begin{figure}[H]
\begin{centering}
\includegraphics[scale=0.45]{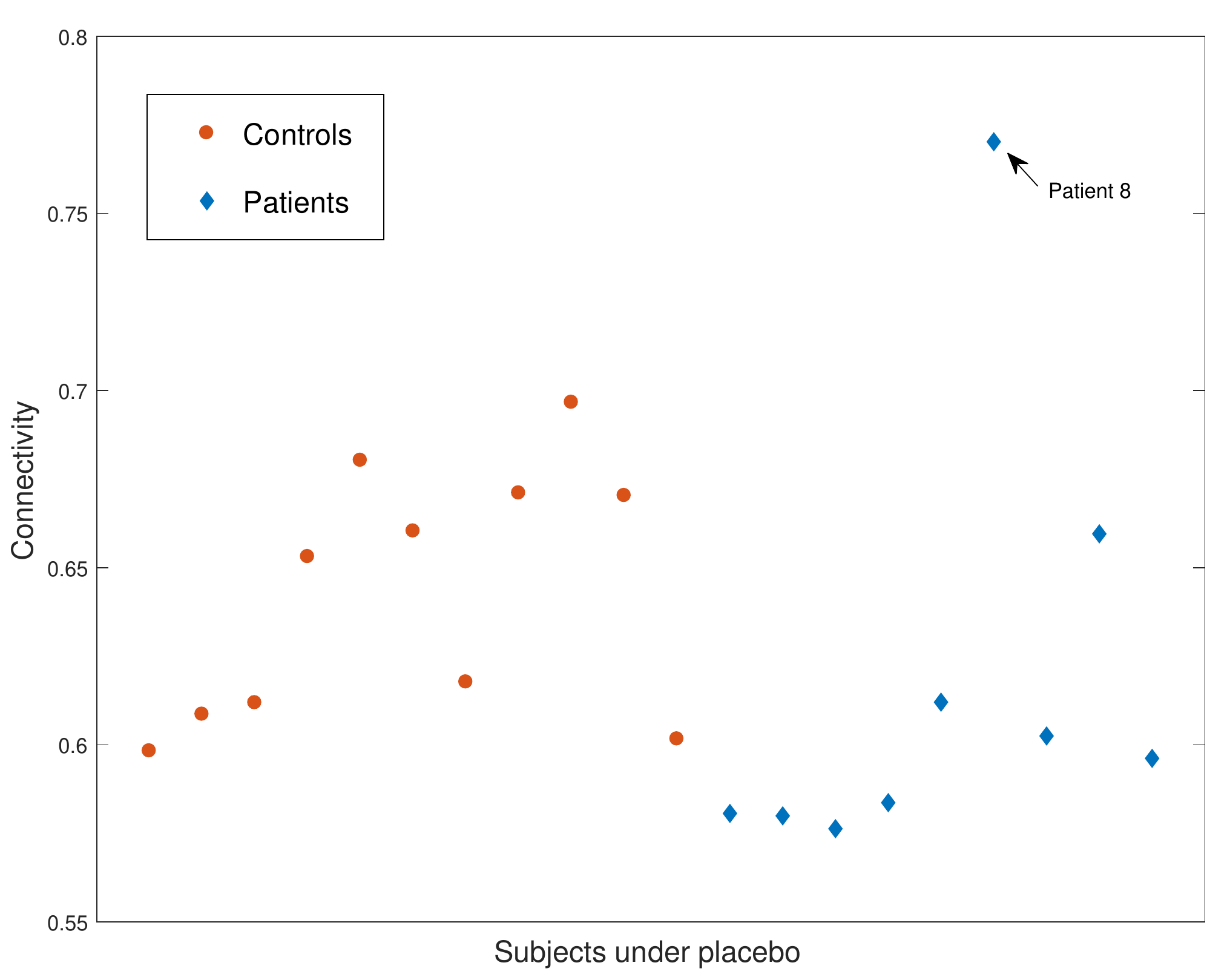}
\par\end{centering}
\protect\caption{\label{fig:ConnectivityPlaecbo} Connectivity for all subjects under Placebo. Note that Patient 8 has an abnormally high connectivity. 
}
\end{figure}

For connectivity, we calculate $\langle W_{ij} \rangle$ for each subject, and we then calculate the means for both controls and patients. We follow the same process for the clustering coefficient. In our preliminary analysis, we explore how these basic quantities differ for different drug treatments. Specifically, we calculate connectivity using the non-thresholded weighted versions of the networks and mean local clustering using the thresholded binary networks. We show our results in Fig.~\ref{fig:clustering}, where for each case we plot the mean and standard deviation across subjects. For each drug treatment, we also perform a two-sample $t$-test on the values of connectivity and mean local clustering coefficients for controls and patients, and we extract a p-value.
We observe small differences in connectivity and mean local clustering coefficients between controls and patients; this difference is smaller than what was reported previously with these data using other approaches~\cite{lynall2010functional}. We also observe that Aripiprazole has a small effect on the connectivity and mean local clustering coefficients of healthy controls but no significant effect on patients, in agreement with other recent work~\cite{EmmaPaper}. Sulpiride appears to have little effect on either group, though we observe a larger difference between controls and patients for mean local clustering coefficient than we do for connectivity. We obtain a p-value of $p \approx 0.0326$ for mean local clustering coefficient and a p-value of $p \approx 0.1680$ for connectivity. We show the connectivity for all subjects under Placebo in Fig.~\ref{fig:ConnectivityPlaecbo}, and we note that Patient 8 has very high connectivity. However, given the sizes of the error bars, we cannot reject the hypotheses that the connectivity and/or mean local clustering coefficients are indistinguishable in the different situations. This suggests that --- at least for this data set --- these simple network diagnostics do not give clear information about whether the drugs have any effects on the structure of functional brain networks. Given the inconclusiveness of these results, we need to do a more sophisticated analysis,

\subsection{Distance Measures} \label{sub:Similarity-and-Distance}

As we mentioned in Section~\ref{sec:Intro}, we aim to classify similar functional brain networks using unsupervised clustering of subjects. A subject is represented by a functional network, so to do this in a systematic way, we define a pairwise distance function between graphs, and we then use this function to compute a distance matrix for a set of subjects. (See step 2 in Fig.~\ref{fig:PaperProcess}.) We consider distance functions based on two rather different aspects of networks.


\subsubsection{Hadamard-like distance} 

One can construct a simple similarity measure between binary networks $\mathbf{A}$ and $\mathbf{B}$ that both have the same number of edges by computing the Hadamard product of the matrices and then summing the entries $A_{ij}B_{ij}$ of the resulting matrix. For binary networks, this sum ($\sum_{i>j}A_{ij}B_{ij}$) is the number of common edges in the networks. Because functional networks are usually thresholded so that one retains only a specified, fixed fractions of edges, we can use this similarity measure to compare adjacency matrices that are extracted from thresholded functional networks. We define the metric
\begin{equation} \label{eq:sim}
	d_1(\mathbf{A},\mathbf{B})=1-\frac{1}{E}\sum_{i>j}A_{ij}B_{ij} \in [0,1] \,,
\end{equation}
which is well-defined when $\mathbf{A}$ and $\mathbf{B}$ have the same number $E$ of edges.

We have proven rigorously (see Appendix \ref{app} for the precise statement of the theorem and proof) that $d_1$ satisfies the properties of a metric. We can thus construct a distance matrix ${\bf D}^{1}$, whose elements ${\bf D}^{1}_{\alpha \beta}$ measure the distance between the functional networks of subjects $\alpha$ and $\beta$. Using ${\bf D}^{1}$ has the advantage of being computationally efficient and based on a mathematically sound metric, although $d_1$ is a rather simplistic measure --- two networks are more distant from each other when they have fewer common edges --- and we do not expect it to capture certain details (e.g., community structure) of the networks.


\subsubsection{Community-structure-based distance} \label{CommunityDistance}

We also use a more sophisticated distance measure, introduced by Onnela et al.~\cite{onnela2012taxonomies}, that is based on network community structure~\cite{Porter2009,Fortunato2016}. It requires using a method of partitioning that assigns each node to a community (i.e., it is a ``hard partition''). In the present paper, we use modularity maximization~\cite{ng2004,newmodlong} and employ the code of Onnela et al., which implements the Louvain method~\cite{blondel}.

Given a network described by its weight matrix $\mathbf{W}$, one can detect communities of this network by maximizing modularity, which one does by minimizing the objective function
\begin{equation} \label{mod}
	\mathcal{H}(\gamma)=-\sum_{i\neq j}\left(W_{ij}-\gamma\frac{r_{i}r_{j}}{2M}\right)\delta(C_{i},C_{j})\,,
\end{equation}
where $\gamma$ is a resolution parameter, $C_{i}$ is the community assignment of node $i$ (and $C_j$ is the community assignment of node $j$), $r_{i}$ is the strength (i.e., sum of incident edge weights) of node $i$, and $M$ is the total edge weight. We consider undirected networks, so we use the Newman--Girvan null-model matrix ${\bf P}$ with elements $P_{ij} = r_{i}r_{j}/(2M)$ \cite{newmodlong,bazzi2016}. For unweighted networks, node strength reduces to degree (i.e., $r_i = k_i$ and $r_j = k_j$), and the total edge weight reduces to the total number of edges (i.e., $M = E$). For each value $\gamma$, minimizing the objective function \eqref{mod} gives a partition of nodes into disjoint communities.
 
Onnela et al. examined so-called ``mesoscopic response functions'' (MRFs) for three quantities that describe, from different perspectives, how a partition of a network changes as a function of $\gamma$. They calculated an effective energy ($\mathcal{H_{\mbox{eff}}}$), an effective entropy ($\mathcal{S_{\mbox{eff}}}$), and an effective number of communities ($\mathcal{\eta_{\mbox{eff}}}$) as functions of a resolution parameter $\xi$ that depends on $\gamma$. As discussed in~\cite{onnela2012taxonomies}, the parameter $\xi$ tracks, in a discrete manner (keeping track of when each effective weight changes sign), which edges have an effective positive weight and which have an effective negative weight when the associated null-model matrix element is taken into account. A given network has a particular profile for $\mathcal{H_{\mbox{eff}}}$, $\mathcal{S_{\mbox{eff}}}$, and $\mathcal{\eta_{\mbox{eff}}}$ (or for any other quantity that one wishes to track~\cite{lee2017}) as a function of $\xi$, and one can then compare a pair of networks based on the differences in these profiles. One can define three distances between a pair of networks, $\alpha$ and $\beta$, as follows:
\begin{align} 
	d_{\alpha \beta}^{\mathcal{H}} &= \intop_{0}^{1}\left|\mathcal{H_{\mbox{eff}}^{\mathrm{\mathit{\alpha}}}}(\xi)-\mathcal{H_{\mbox{eff}}^{\mbox{\ensuremath{\beta}}}}(\xi)\right|d\xi\,,  \label{trio1} \\
	d_{\alpha \beta}^{\mathcal{S}} &= \intop_{0}^{1}\left|\mathcal{S_{\mbox{eff}}^{\mathrm{\mathit{\alpha }}}}(\xi)-\mathcal{S_{\mbox{eff}}^{\mbox{\ensuremath{\beta}}}}(\xi)\right|d\xi\,, \label{trio2} \\
	d_{\alpha \beta}^{\mathcal{\eta}} &= \intop_{0}^{1}\left|\mathcal{\eta_{\mbox{eff}}^{\mathrm{\mathit{\alpha}}}}(\xi)-\mathcal{\eta_{\mbox{eff}}^{\mbox{\ensuremath{\beta}}}}(\xi)\right|d\xi\,. \label{trio3}
\end{align}

The trio of distances in Eqs.~\eqref{trio1}--\eqref{trio3} capture different aspects of community structure. The effective energy ($\mathcal{H_{\mbox{eff}}}$) is a rescaled version of the objective function $\mathcal{H}$, the effective entropy ($\mathcal{S_{\mbox{eff}}}$) represents the level of homogeneity in the sizes of the detected communities, and the effective number of communities ($\mathcal{\eta_{\mbox{eff}}}$) is a rescaled version (with respect to network size) of the total number of communities. From these distances matrices, we construct a single distance matrix (although one can also separately study distance matrices constructed using these, or other, distances). To construct this distance matrix, we project each 3-dimensional coordinate using principal component analysis (PCA) and keep the first component. In other words, we construct a distance matrix by calculating a linear combination of the three distance measures:
\begin{equation}\label{mrfd}
	d_{\alpha \beta}^P=w_{\mathcal{H}}d_{\alpha \beta}^{\mathcal{H}} + w_{\mathcal{S}}d_{\alpha \beta}^{\mathcal{S}} + w_{\eta}d_{\alpha \beta}^{\eta}\,,
\end{equation}	
where the weights $w_{\ell}$ (with $\ell \in \{\mathcal{H},\mathcal{S},\eta\}$) are the coefficients of the first principal component. There are a total of 60 networks (11 controls and 9 patients, each of which is on 3 different drug treatments). We calculate the matrix composed of $60 \times 59 / 2$ (the total number of network pairs) rows and $3$ columns, where each column corresponds to the vector representation of the upper triangle of one of the distance matrices $\mathbf{D^{\mathcal{H}}}$, $\mathbf{D^{\mathrm{\mathsf{\mathit{S}}}}}$, $\mathbf{D^{\eta}}$. We perform a PCA on this matrix to create a distance matrix $\mathbf{D}^{P}$. 

The final outcome of the above calculation is a $60\times60$ distance matrix $\mathbf{D}^{P}$, where each entry measures the distance between networks $\alpha$ and $\beta$ based on how the community structure of each network varies as a function of the resolution parameter $\xi$. We henceforth use the term ``MRF distance'' for the quantity that we compute in Eq.~\eqref{mrfd}.


\subsection{Hierarchical Clustering} \label{sub:Clustering}

Once we have our distance matrix (see Section~\ref{CommunityDistance}), we take a submatrix of it for each of the comparisons in Fig.~\ref{fig:Comparisons-1}. For example, if we are comparing controls and patients under the drug Aripiprazole, we keep only the rows and columns that correspond to this drug, leaving us with a $20 \times 20$ distance matrix, where the rows and columns correspond to the $11$ controls and $9$ patients. We then cluster the new, smaller distance matrix using one of numerous possible methods. For simplicity, we use average linkage clustering to group similar subjects (i.e., similar networks) together and show our results in the form of dendrograms. We then order the leaves of the dendrogram to maximize the sum of the similarities between adjacent leaves by reordering its branches (without further partitioning clusters). We color the leaves of the dendrograms based on their annotations: patients or controls without drugs, patients or controls on one drug, or patients and controls on the other drug. 


\section{Main Results} \label{sec:Results}

\begin{figure}
\begin{centering}
\includegraphics[scale=0.6]{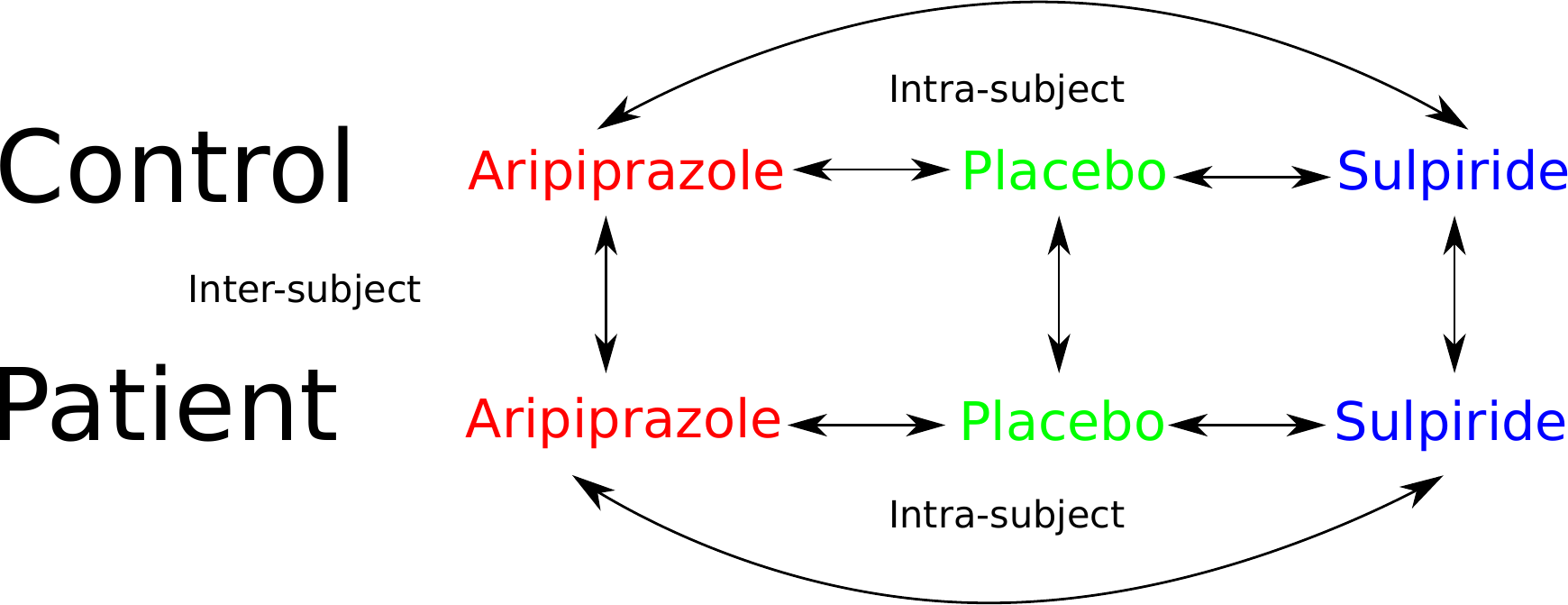}
\par\end{centering}
\protect\caption{\label{fig:Comparisons-1} 
Illustration of possible comparisons between the groups of subjects and different drug treatments. 
}
\end{figure}

As we mentioned in Section~\ref{sec:Intro} and depicted in Fig.~\ref{fig:Comparisons-1}, we are going to make a total of 9 comparisons, including both inter-subject ones (different groups under the effect of the same drug) and intra-subject ones (the same group under the effect of different drugs). In our ensuing discussions, we present the results of these comparisons.

\subsection{Inter-subject Comparisons} \label{sec:InterSubject}

\begin{figure}
\centering{}
\includegraphics[scale=0.6]{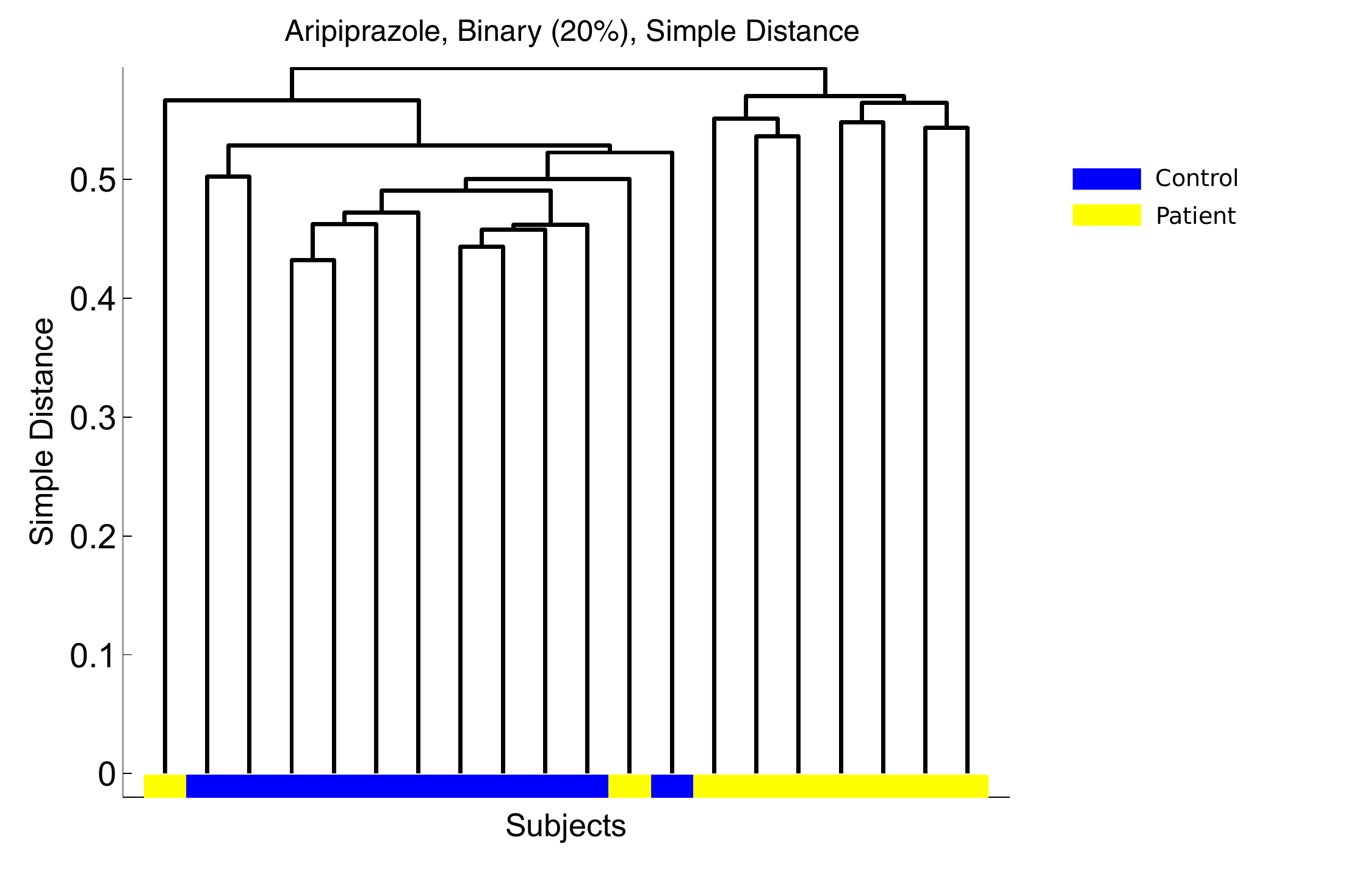}
\protect\caption{\label{fig:Dendrogram-for-thesimple}Dendrogram for the drug Aripiprazole in which we compare the 11 controls and 9 patients using the distance measure $\text{d}_1({\mathbf{A},\mathbf{B}})$. There is some separation between patients and controls.}
\end{figure}

We do inter-subject comparisons using the procedure that we outlined in Fig.~\ref{fig:PaperProcess}. We start by comparing controls and patients under the effects of the drug Aripiprazole using the simple distance measure $\text{d}_1({\mathbf{A},\mathbf{B}})$ from Eq.~\eqref{eq:sim}. We show the resulting dendrogram in Fig.~\ref{fig:Dendrogram-for-thesimple}. We observe some separation between patients and controls.

\begin{figure}
\begin{centering}
\includegraphics[scale=0.6]{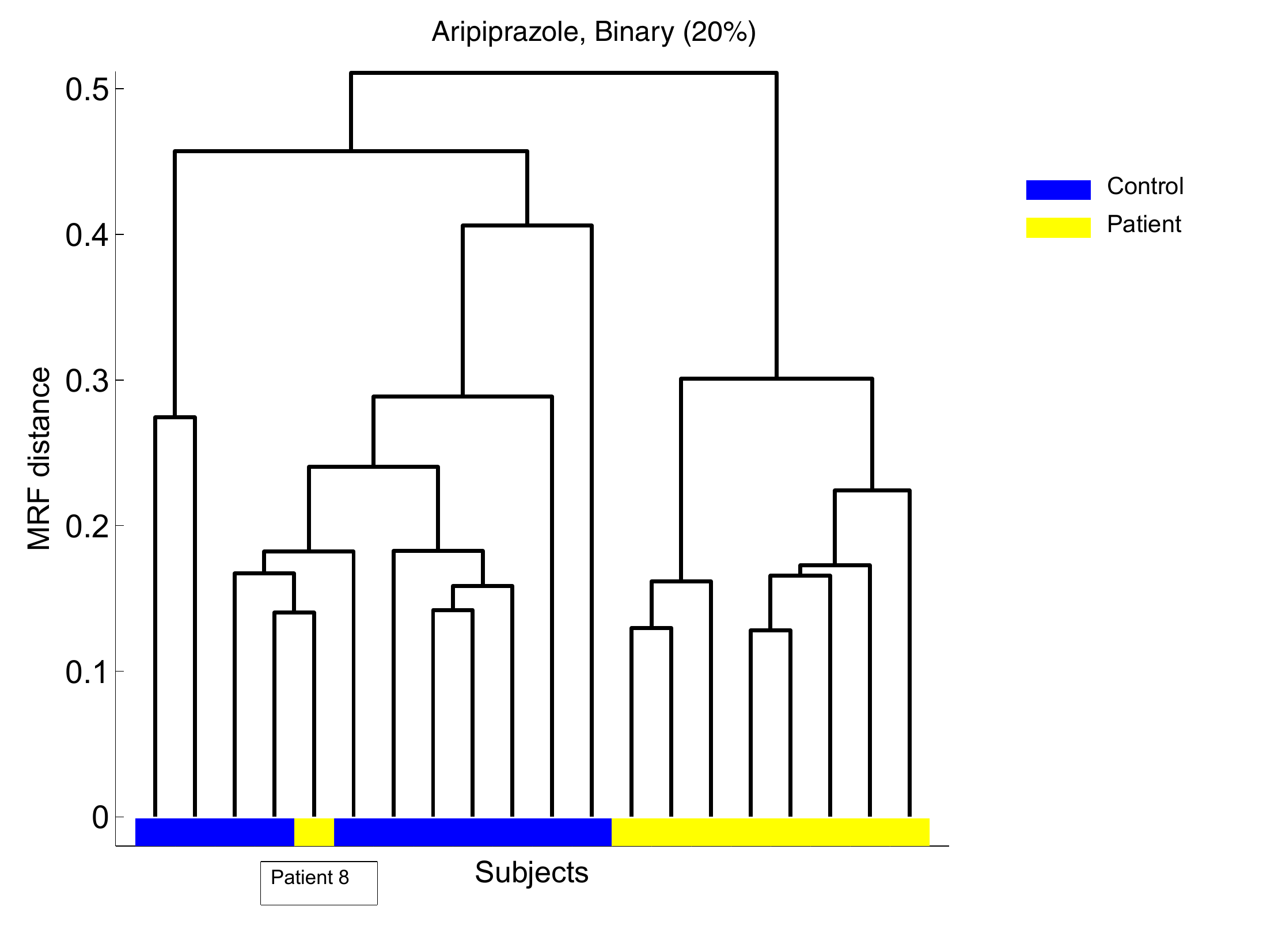}
\par\end{centering}
\protect\caption{\label{fig:Aripiprazole}Dendrogram for the MRF analysis of functional brain networks for the drug Aripiprazole. We compare the 11 controls and 9 patients using the distance measure $\mathbf{D^{\mathrm{\mathit{p}}}}$. There is a clear separation between patients and controls, although Patient 8 appears with the control group.}
\end{figure}

To do a more sophisticated analysis, we then compute a dendrogram on the same data using the MRF distance matrix $\mathbf{D}^{P}$ (see Section~\ref{sub:Similarity-and-Distance}). We show the resulting dendrogram in Fig.~\ref{fig:Aripiprazole}. The separation between patients and controls is now better, and we correctly classify almost every individual. The only exception is Patient 8, who is assigned to the same group as the controls. Although this misclassification seems surprising at first, it agrees with our previous calculations (see Fig.~\ref{fig:ConnectivityPlaecbo}), which also suggest that Patient 8 has different network characteristics than the other patients.

The above result suggests that, under the drug Aripiprazole, we are able to almost completely distinguish patients from controls, based only on information about their community structure. This also suggests that the distance matrix $\mathbf{D}^{P}$ incorporates more meaningful information than the simplistic distance measure in Eq.~\eqref{eq:sim}, so we use only the former for our subsequent computations.

\begin{figure}
\begin{minipage}[t]{0.45\textwidth}%
\begin{center}
\includegraphics[scale=0.4]{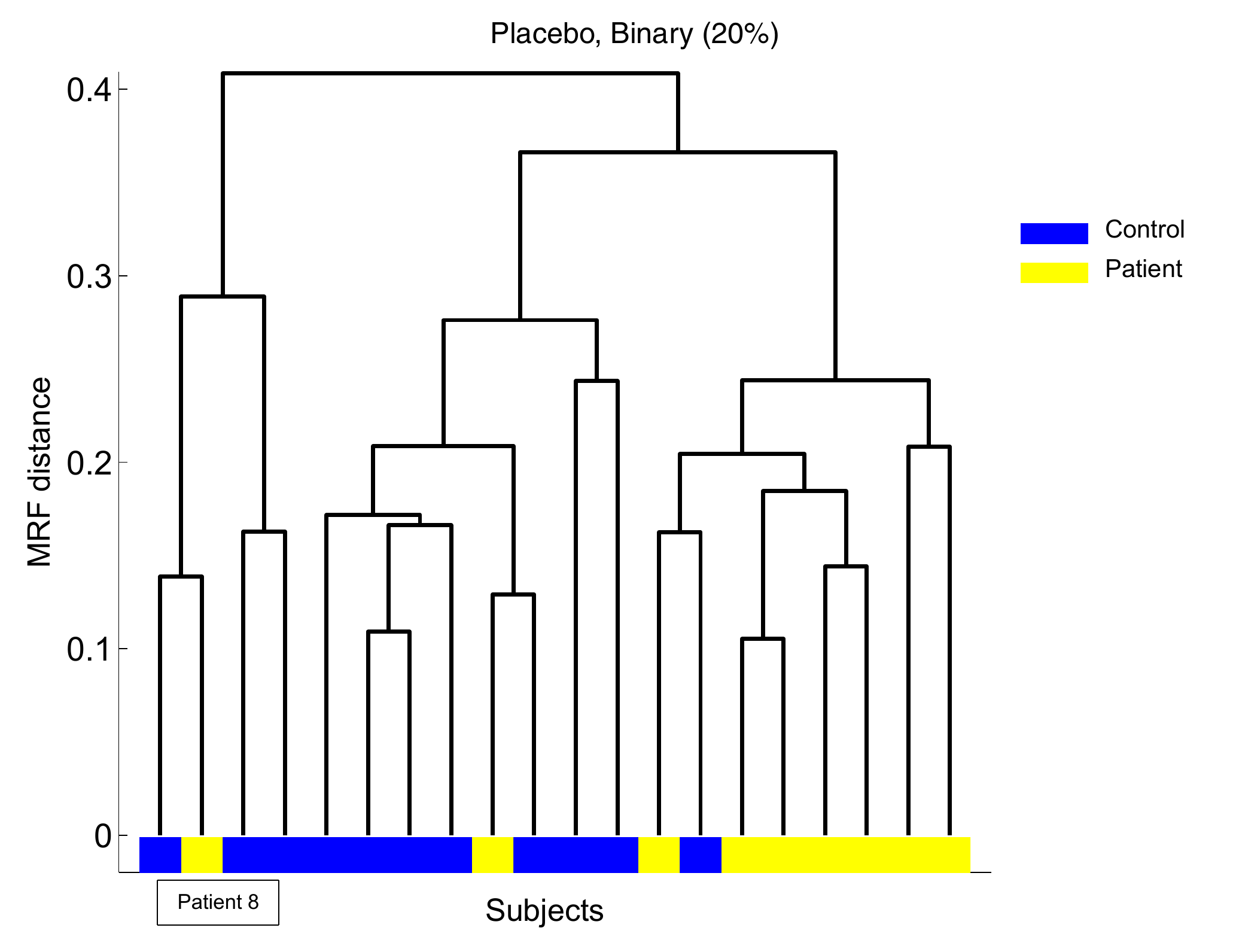}
\par\end{center}%
\end{minipage}\hfill{}%
\begin{minipage}[t]{0.45\textwidth}%
\begin{center}
\includegraphics[scale=0.4]{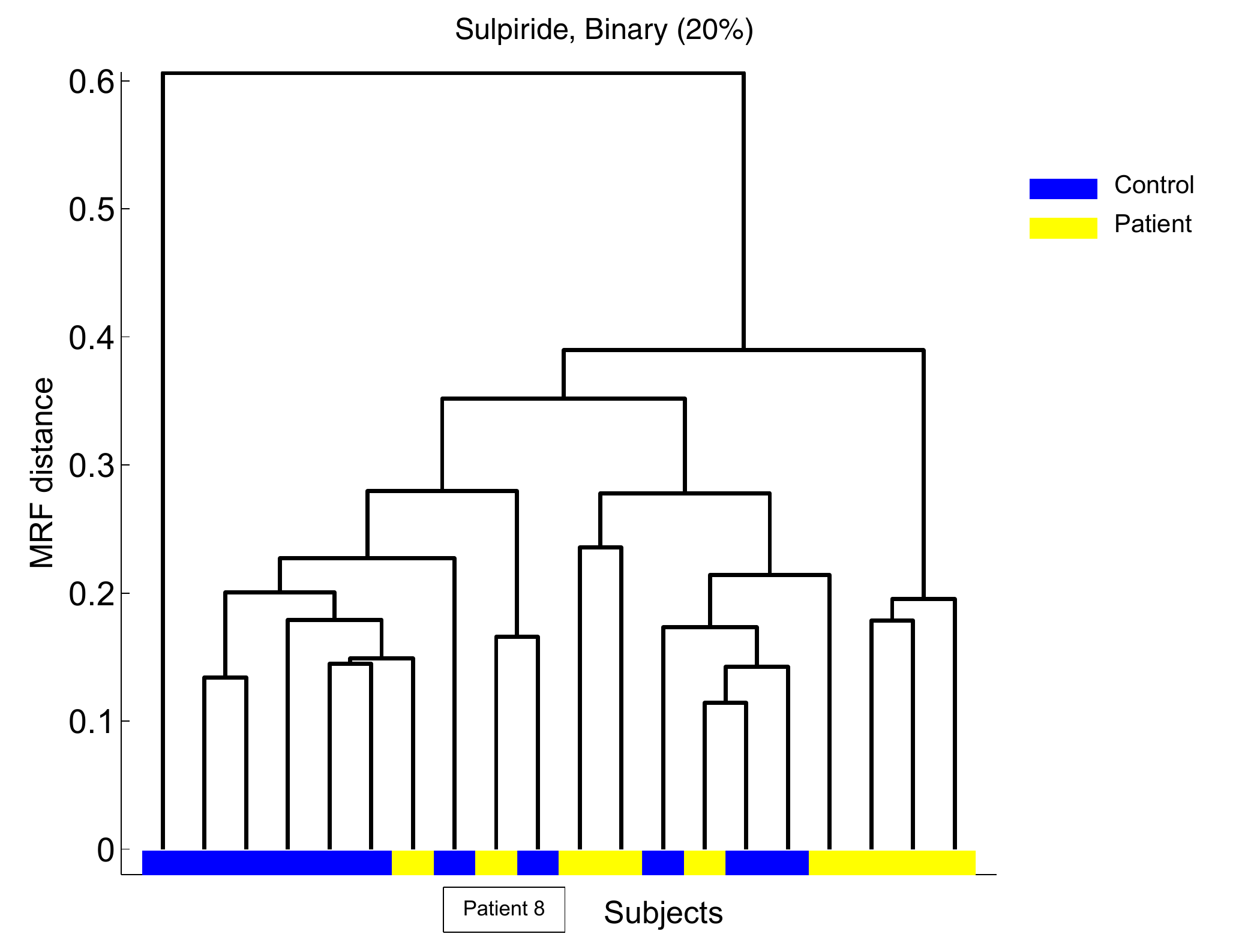}
\par\end{center}%
\end{minipage}\protect\caption{\label{fig:PlaSul}Dendrogram for our MRF analysis of functional brain networks for (left) Placebo and (right) the drug Sulpiride. In order of most successful to least successful (compare this figure to Fig.~\ref{fig:Aripiprazole}), the clustering performs best for Aripiprazole, second-best for Placebo, and worst for Sulpiride. 
}
\end{figure}

We show the analogous results comparing controls and patients under Placebo in the left panel of Fig.~\ref{fig:PlaSul}. In this case, we still observe a relatively good separation between patients and controls, in agreement with previous results that functional networks encode biomarkers that separate patients diagnosed with schizophrenia from healthy controls~\cite{lynall2010functional, zalesky2012relationship}. In this situation too, Patient 8 appears to be more similar to the controls than to the other patients. Even more interesting, we observe a {\it less-clear separation} between the controls and patients than we did under Aripiprazole. We thus conclude that Aripiprazole alters community structure for at least one group and that this alteration makes it easier to distinguish the patient and control groups. However, it is not obvious whether Aripiprazole is affecting the structure of the functional brain networks of patients, controls, or both.

In the right panel of Fig.~\ref{fig:PlaSul}, we show our results for computations of functional brain networks for individuals under the influence of Sulpiride. The control and patient groups are now less distinct from each other than they were with Placebo. This suggests that Sulpiride has a mild but detectable effect of increasing the similarity between the community structures of patients and controls. Again, it is not clear whether Sulpiride affects the functional brain networks of patients, controls, or both.

\subsection{Intra-subject Comparisons}\label{intra}

To examine the effects of the drug treatments on network structure, we make intra-subject comparisons, such as comparing the control group under Aripiprazole to the control group under Sulpiride. We do these comparisons using the procedure that we outlined in Fig.~\ref{fig:PaperProcess}. 


\subsubsection{Aripiprazole versus Placebo}\label{Aripiprazole}

For our intra-subject comparisons (see Fig.~\ref{fig:Comparisons-1}), we first compare the effect of Aripiprazole on the functional brain networks of controls to those of patients. To do this, we use all 11 controls under Aripiprazole and the same 11 controls under Placebo and do average linkage clustering on the associated $22 \times 22$ distance matrix with MRF distances. We also do average linkage clustering using the MRF distance for the $18 \times 18$ distance matrix that we obtain by considering the 9 patients under Aripiprazole and the same patients under Placebo.

\begin{figure}
\begin{centering}
\includegraphics[scale=0.6]{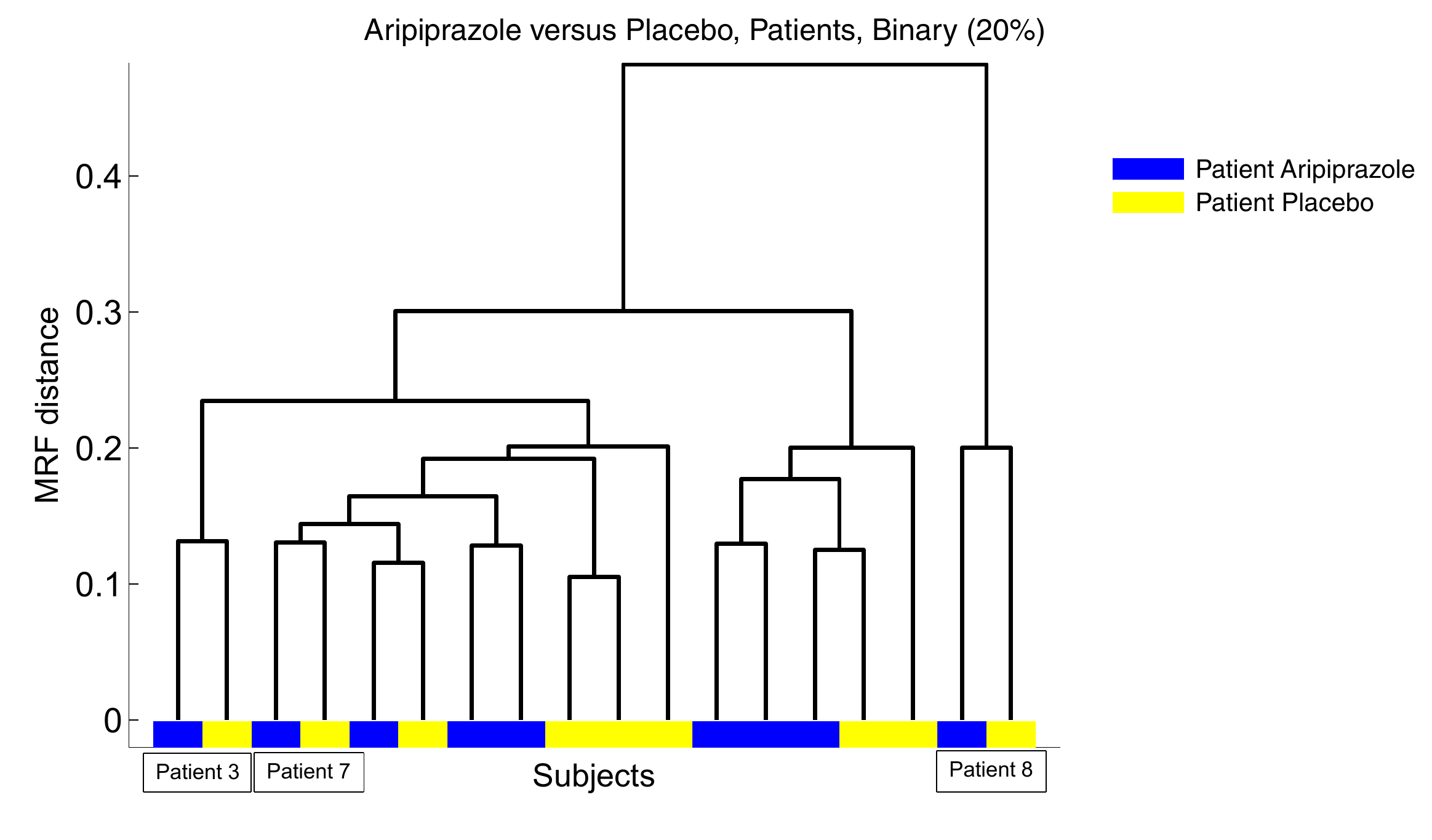}
\par\end{centering}
\protect\caption{\label{fig:AvsPPatient}
Dendrogram for our MRF analysis of functional brain networks for our comparison between Aripiprazole and Placebo for the patient group. Each patient thus appears twice on the horizontal axis. There is no clear separation between the two drugs, and some patients (e.g., 3, 7, and 8) cluster very close to themselves, suggesting there there is very little difference in community structure in the networks under Placebo and under Aripiprazole in these patients.
}
\end{figure}

\begin{figure}
\begin{centering}
\includegraphics[scale=0.58]{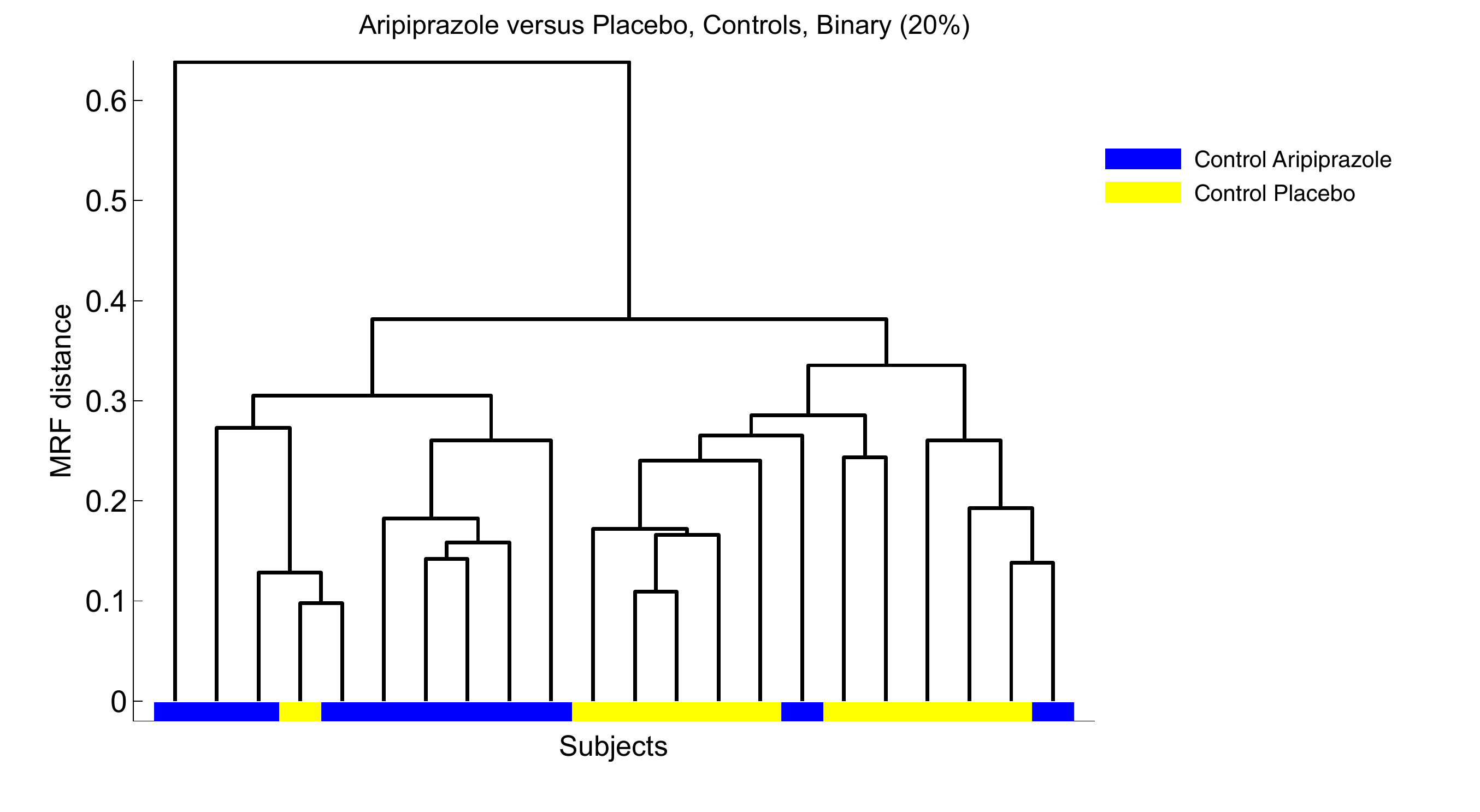}
\par\end{centering}
\protect\caption{\label{fig:AvsPControl} Dendrogram for our MRF analysis of functional brain networks for our comparison between Aripiprazole and Placebo for the control group. We observe a clear separation between networks under the two drug treatments.
}
\end{figure}

In Fig.~\ref{fig:AvsPPatient}, we show the dendrogram for our comparison between Aripiprazole and Placebo for patients. At the coarsest level of detail (i.e., a separation for a large MRF distance in the dendrogram), we observe that both the Aripiprazole and Placebo network of Patient 8 is grouped away from those of the other patients. This is consistent with our prior results: we saw in Fig.~\ref{fig:ConnectivityPlaecbo} that Patient 8 has a much higher connectivity than the other patients and saw in Fig.~\ref{fig:Aripiprazole} that Patient 8 was grouped with the controls. At the finest level of detail, we also find for both Aripiprazole and Placebo that Patients 3 and 7 cluster close to each other. This suggests there is little community structure in these patients under Aripiprazole compared to a Placebo. We thus expect, given the inter-subject comparisons in Section~\ref{sec:InterSubject}, that Aripiprazole does affect community structure in controls. We confirm this hypothesis in Fig.~\ref{fig:AvsPControl}, where we observe that controls under Aripiprazole are clearly separated from controls under Placebo.

\subsubsection{Sulpiride versus Placebo} 

\begin{figure}
\begin{centering}
\includegraphics[scale=0.6]{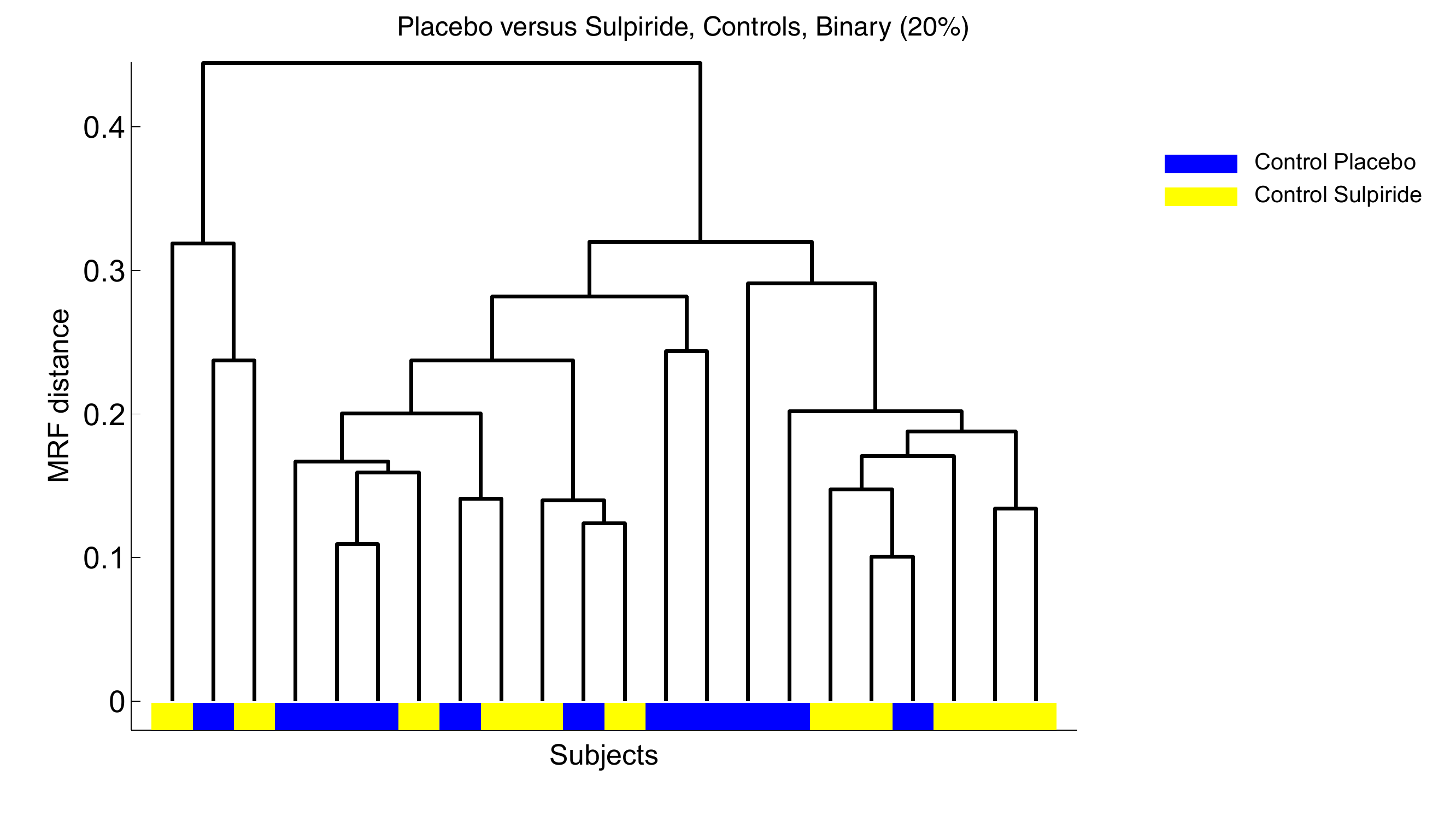}
\par\end{centering}
\protect\caption{\label{fig:PvsSControls}Dendrogram for our MRF analysis of functional brain networks for our comparison between Sulpiride and Placebo for the control group.}
\end{figure}

\begin{figure}
\begin{centering}
\includegraphics[scale=0.6]{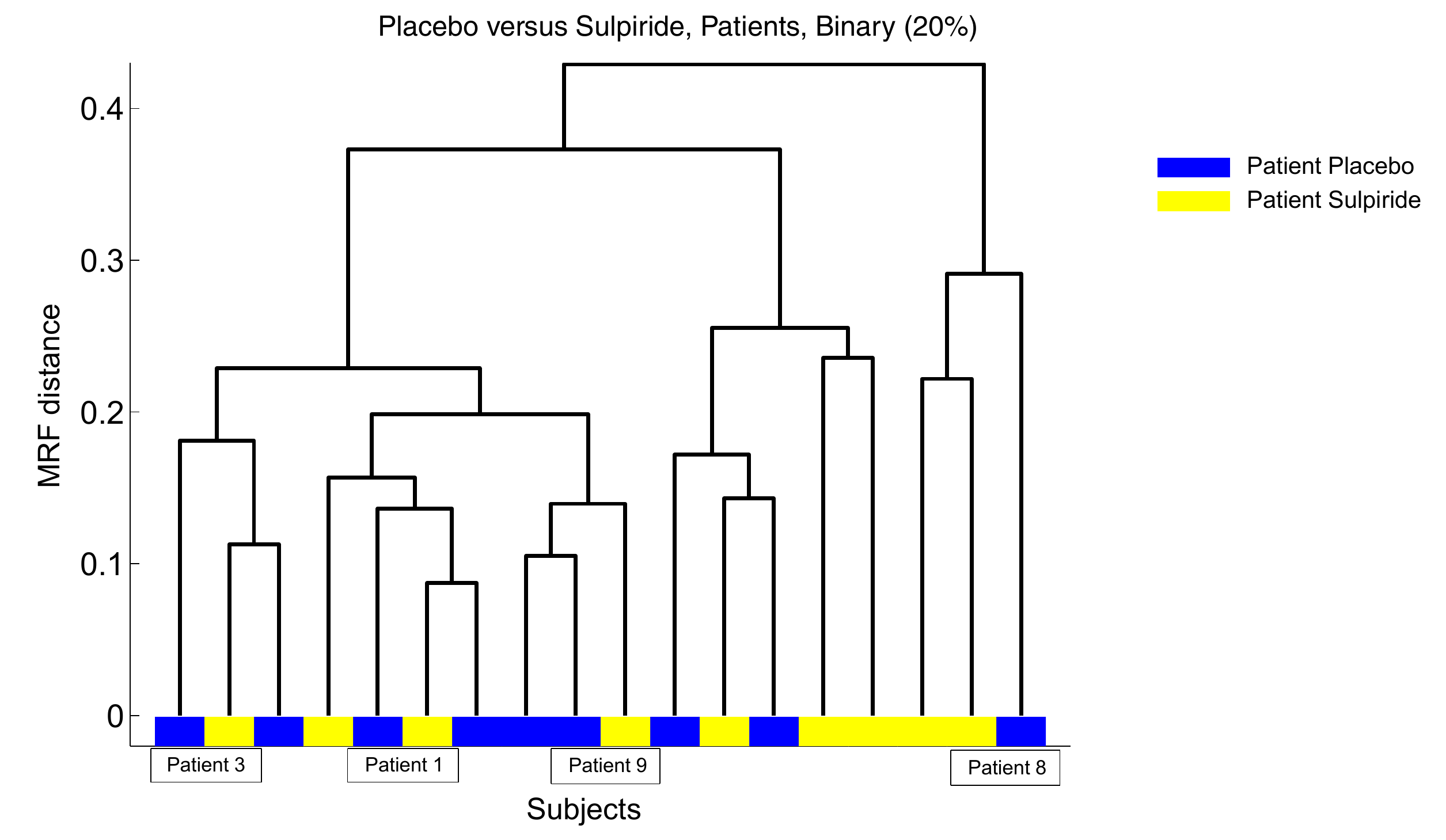}
\par\end{centering}
\protect\caption{\label{fig:PvsSPatientsB}Dendrogram for our MRF analysis of functional brain networks for our comparison between Sulpiride and Placebo for the patient group. As with our comparison of Placebo to Aripiprazole, identical patients appear close together, and Patient 8 is again distant from the others.
}
\end{figure}

In Section~\ref{Aripiprazole}, we observed a very clear separation between controls and patients under the drug Aripiprazole and evidence (though the situation is less clear) of separation under Placebo. We observed an even lesser separation in Sulpiride. We hypothesized that Sulpiride has a mild but detectable effect of increasing the similarity between community structure in patients and controls, and we therefore hypothesize that Sulpiride affects community structure of either patients or controls (or both), in agreement with~\cite{EmmaPaper}. In Fig.~\ref{fig:PvsSControls}, we show a dendrogram of the intra-subject comparison of Placebo versus Sulpiride in controls. We do not observe any clear clustering. We also do not observe any clustering in the same comparison for patients (see Fig.~\ref{fig:PvsSPatientsB}). We therefore do not find any clear indication of why Sulpiride seems to make controls and patients less distinguishable from each other. Additionally, we do not observe a clear separation under Placebo or under Sulpiride either for controls (see Fig.~\ref{fig:PvsSControls}) or for patients (see Fig.~\ref{fig:PvsSPatientsB}).

\subsubsection{Aripiprazole versus Sulpiride}

\begin{figure}
\begin{centering}
\includegraphics[scale=0.6]{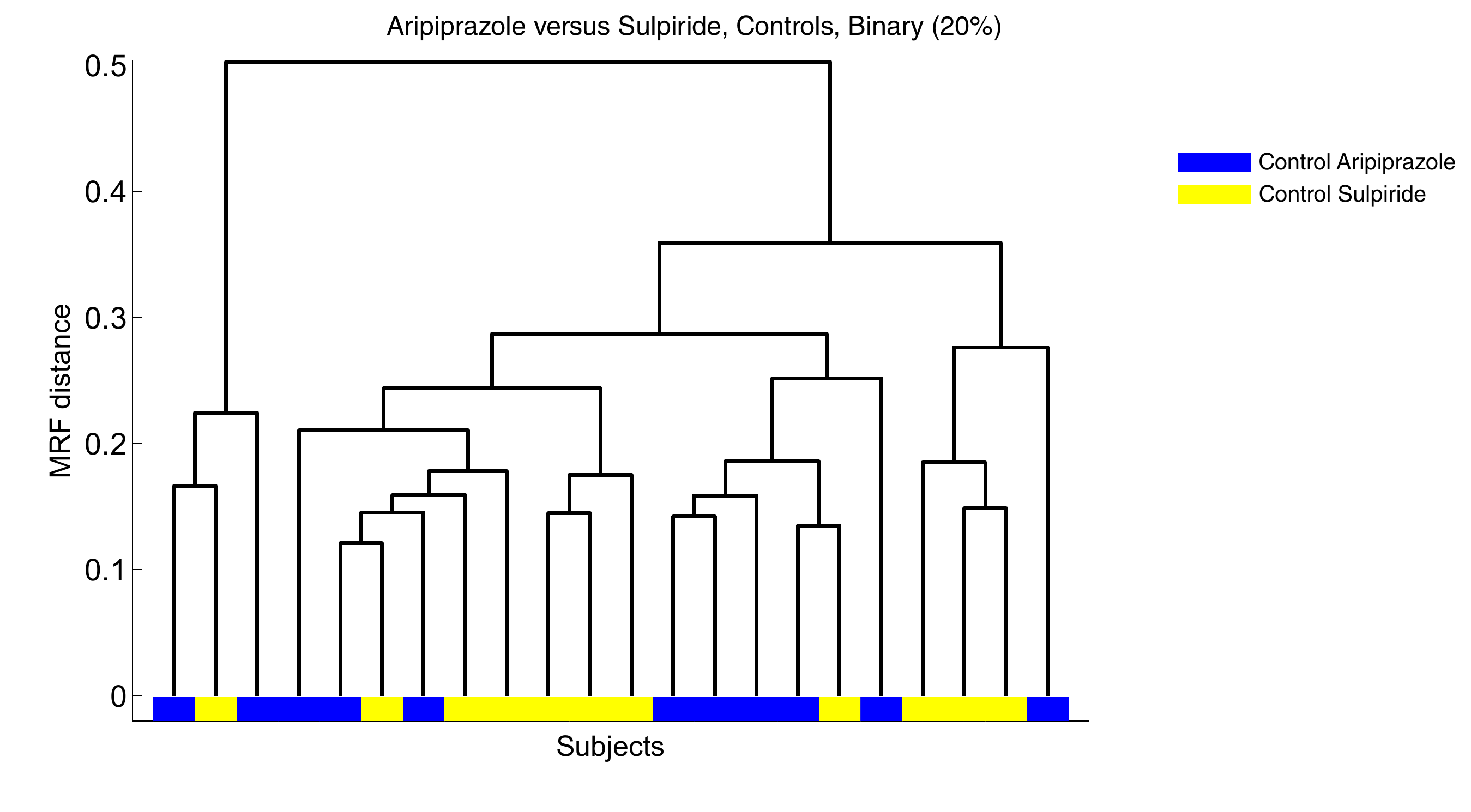}
\par\end{centering}
\protect\caption{\label{fig:AvsSControls}Dendrogram for our MRF analysis of functional brain networks for our comparison between Aripiprazole and Sulpiride for the control group.}
\end{figure}

\begin{figure}
\begin{centering}
\includegraphics[scale=0.6]{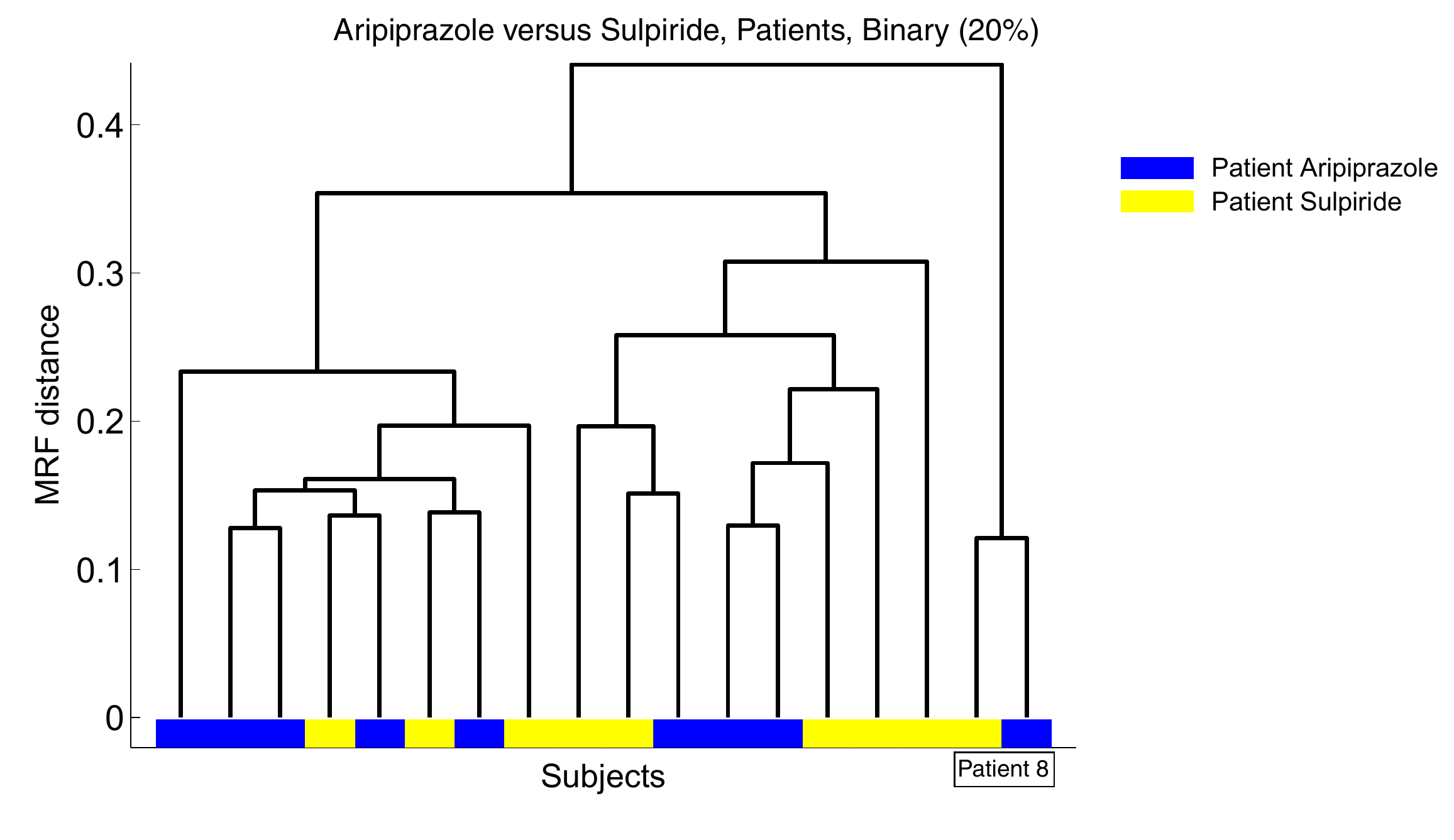}
\par\end{centering}
\protect\caption{\label{fig:AvsSPatients}Dendrogram for our MRF analysis of functional brain networks for our comparison between Aripiprazole and Sulpiride. In both the Aripiprazole and Sulpiride networks, it is once again easy to distinguish Patient 8 from the other patients.
}
\end{figure}

We can partly distinguish controls under Aripiprazole versus Sulpiride (see Fig.~\ref{fig:AvsSControls}). This is unsurprising, given that we found (see Section~\ref{Aripiprazole}) that Aripiprazole alters community structure in controls. We do not observe any obvious difference for patients under Aripiprazole versus Sulpiride (see Fig.~\ref{fig:AvsSPatients}).

\subsection{Synthesis of our Results from Hierarchical Clustering}

Our results from average linkage clustering of collections of functional brain networks using the distance functions yield the following conclusions:
\begin{itemize}
\item{Aripiprazole affects community structure in controls, but not in patients; and it thereby facilitates the distinction between controls and patients under the effect of this drug treatment.}
\item{Sulpiride reduces the distinguishability between patients and controls, though our intra-subject computations were inconclusive in both patients and controls.}
\end{itemize}


\section{Conclusions and Discussion} \label{sec:Discussion}

We used network analysis to examine the effects of two therapeutic antipsychotics --- Aripiprazole and Sulpiride --- on the structure of functional brain networks of both healthy controls and patients who have been diagnosed with schizophrenia. Using mesoscopic response functions, we compared community structures of functional brain networks of these individuals under the effects of Aripiprazole, Sulpiride, and a placebo.

We will now summarize the results of our computations. However, before doing so, we stress that when interpreting the results of fMRI studies, it is very important to consider the cautionary notes in~\cite{eklund2016}. These complications notwithstanding, our computations produced several interesting results. First, we did a reasonable job of distinguishing between controls and patients under Placebo, and we did a much better job of distinguishing the two groups under Aripiprazole. This suggests that Aripiprazole has a larger effect on community structure in one of the two groups than in the other. By comparing controls under Aripiprazole and under Placebo, we concluded that Aripiprazole appears to improve the distinguishability between patients and controls primarily through its effects on community structure in controls (i.e., healthy individuals). Our observations that Aripiprazole primarily affects community structure in controls is consistent with the results of~\cite{EmmaPaper}, who reported that Aripiprazole has a radical effect on the organization of healthy brain networks but decreases the performance of healthy individuals at cognitive tasks.

Our results for the drug Sulpiride are mixed. We found that under Sulpiride, patients are slightly closer to controls than they are under Aripiprazole or Placebo. This is also consistent with~\cite{EmmaPaper}, and it suggests that Sulpiride has a mild effect on community structure that is appreciably larger than, for instance, the effect of Aripiprazole on community structure in patients (which we observed to be very small). We have not been able to clearly establish the origin of this effect, as our intra-group comparisons suggest that community structure in both controls and patients is mostly unaltered by Sulpiride.

Mesoscale network structures such as community structure are well-known to be important for functional brain networks \cite{Betzel2016}, and network analysis more generally is a useful approach for disentangling structure, function, and their complex interrelations in the brain. In the present paper, we used community structure and mesoscopic response functions for a classification task in time-independent, monolayer functional brain networks. Extending these results and analysis to time-dependent and multilayer settings \cite{kopell2014,buldu2017} is an interesting open problem.


\section*{Acknowledgements}

LL acknowledges funding from EPSRC Early Career Fellowship (EP/P01660X/1), and RF acknowledges funding from the EPSRC (through DTP EP/N013492/1). SHL was supported by a Gyeongnam National University of Science and Technology Grant in 2018--2019. We thank Sebastian Ahnert for helpful discussions, and Ameera Patel for discussions about pre-processing the raw fMRI data. We also thank Ulrich M\"uller for recruiting and scanning participants in the original experimental work, and we thank Melissa Lever for early computational work on this project. Data collection was supported by a grant from Bristol Myers Squibb to Robert Kerwin at King's College London.


\begin{appendix}


\section{Appendix: Metric properties of $d_1$}\label{app}

In this appendix, we state and prove a theorem on the metric properties of $d_1$ (which we defined in Eq.~\eqref{eq:sim}) that is slightly more general than the one that we used in the main text. The result in the main text follows from it as a trivial corollary.

\begin{theorem}\label{thm1}
{\it Let ${\cal S}_n(E)$ be the set of $n \times n$ square matrices with entries of $0$ or $1$ (i.e., ``binary matrices''), where the number $E$ of $1$ entries satisfies $E<n^2$. Let ${\bf A}\,,{\bf B}\in {\cal S}_n(E)$ be two arbitrary elements of the set. Consider the function $d_1$ defined by
\begin{equation}\label{above}
	d_1:{\cal S}_n(E)\times {\cal S}_n(E)\to [0,1]\,,\quad   d_1(A,B)=1-\frac{1}{E}\sum_{i=1}^n\sum_{j=1}^nA_{ij}B_{ij}\,.
\end{equation}	
The function $d_1$ is a metric.}
\end{theorem}

The definition in \eqref{above} for $d_1$ is slightly more general than the one in Eq.~\ref{eq:sim}, as here we are not assuming that (1) ${\bf A}$ and ${\bf B}$ are symmetric or that (2) there are no $1$ entries in the diagonal (so $E$ is the number of $1$ entries). In the main text, we imposed some restrictions on $E$ that are not present here: we used $E$ to denote the number of edges in an associated network, so for Eq.~\ref{eq:sim} (which is designed to deal with unweighted, undirected adjacency matrices with no self-loops), one needs either to restrict to the case in which there are no $1$ entries in the main diagonal and then do the relabeling $E\to E/2$ or to relabel the summation indices with $\sum_{i=1}^n\sum_{j=1}^n\to \sum_{i>j}$. The theorem that we stated in the main text is thus a special case of Theorem \ref{thm1}.

\begin{proof}
To prove that $d_1$ is a metric, we need to prove four properties: nonnegativity, identity of indiscernibles, symmetry, and the triangle inequality. The first three properties are satisfied trivially:
\begin{enumerate}
\item Nonnegativity: By construction, $\sum_{i=1}^n\sum_{j=1}^nA_{ij}B_{ij} \leq E$, so $d_1(\mathbf{A},\mathbf{B})\geq 0$.
\item Identity of indiscernibles: $d_1(\mathbf{A},\mathbf{B})=0 \Leftrightarrow \sum_{i=1}^n\sum_{j=1}^nA_{ij}B_{ij}=E$. However, by definition, the matrices are binary and have $E$ entries with the value $1$, so $\sum_{i=1}^n\sum_{j=1}^nA_{ij}B_{ij}=E \Leftrightarrow  \mathbf{A} = \mathbf{B}$.
\item Symmetry: This arises trivially from the commutative property of the scalar product: $A_{ij}B_{ij}=B_{ij}A_{ij}$.
\end{enumerate}
To prove the fourth property (the triangle inequality), we need to show that 
\begin{equation}\label{triangle}
	\text{for all } \,{\bf A}, {\bf B}, {\bf C} \in {\cal S}_n(E)\,, \quad d_1(\mathbf{A},\mathbf{B}) + d_1(\mathbf{B},\mathbf{C}) \geq d_1(\mathbf{A}\,.\mathbf{C})\,.
\end{equation} 
This part is more subtle, and we need to break the proof into a couple of steps. We start by defining a \emph{matrix $\delta$-perturbation}.

\begin{mydef}({\sc Matrix $\delta$-perturbation}) Let ${\bf A} \in{\cal S}_n(E)$, and let $\delta$ be a positive integer such that $0<\delta<E$. The matrix $\tilde{{\bf A}}^{(\delta)}$ is a \underline{$\delta$-perturbation} of {\bf A} if $\tilde{{\bf A}}^{(\delta)}$ is constructed by taking {\bf A} and changing the position of $\delta$ of the $1$ entries.  
\end{mydef} 

To illustrate this definition, we show an example of a matrix and a 1-perturbation of that matrix in $S_3(3)$:
\begin{equation} 
{\bf Z}=
\begin{bmatrix}
    1       & 0 & 0 \\
    1       & 0 & 0  \\
    0       & 1 & 0 
\end{bmatrix}\,, \quad \tilde{{\bf Z}}^{(1)}=
\begin{bmatrix}
    0       & 0 & 1 \\
    1       & 0 & 0  \\
    0       & 1 & 0 
\end{bmatrix}\,.
\end{equation}

It is clearly the case that $\tilde{{\bf A}}^{(\delta)}\in {\cal S}_n(E)$. It is also true that
\begin{equation*}
	\sum_{i=1}^n\sum_{j=1}^nA_{ij}\tilde{A}_{ij}^{(\delta)}=E-\delta \Rightarrow d_1(\tilde{{\bf A}}^{(\delta)},{\bf A})=\delta/E\,.
\end{equation*}	 
Starting from an arbitrary element of ${\cal S}_n(E)$, one can reach any other element by applying an appropriate $\delta$-perturbation.  Therefore, equipped with the $\delta$-perturbation, ${\cal S}_n(E)$ is a unary system. This property is important for guaranteeing completeness.

To prove Eq.~\eqref{triangle}, it is equivalent to prove that 
\begin{equation*}
	\text{for all } \, {\bf A}, {\bf B}, {\bf C} \in {\cal S}_n(E)\,, \quad {\cal X}:=\sum_{i=1}^n\sum_{j=1}^n(A_{ij}B_{ij}+B_{ij}C_{ij}-A_{ij}C_{ij})\leq E\,.
\end{equation*}
We are ready to prove this latter inequality. We start with a degenerate case. Consider an arbitrary ${\bf A} \in {\cal S}_n(E)$ and set ${\bf A}={\bf B}={\bf C}$; in this case, ${\cal X}=\sum_{i=1}^n\sum_{j=1}^nA_{ij}A_{ij}=E\leq E$.

To generate all possible triples of matrices $\{ {\bf A}, {\bf B}, {\bf C}\}$, without loss of generality, we now consider an arbitrary (but fixed) ${\bf A} \in {\cal S}_n(E)$; and we use $\delta$-perturbations to generate all instances of ${\bf B}$ and ${\bf C}$. That is,
\begin{equation*}
	{\bf B}:= \tilde{{\bf A}}^{(\delta_b)}, \ {\bf C}:= \tilde{{\bf A}}^{(\delta_c)}, \ \delta_b,\delta_c\geq 0\,.
\end{equation*}	
All possible triples can be expressed in this form.

Let's evaluate $\cal X$. The first term is 
\begin{equation*}
	\sum_{i=1}^n\sum_{j=1}^nA_{ij}B_{ij}=\sum_{i=1}^n\sum_{j=1}^nA_{ij}\tilde{A}_{ij}^{(\delta_b)}=E-\delta_b\,;
\end{equation*}	
the second term is
\begin{equation*}
	\sum_{i=1}^n\sum_{j=1}^nB_{ij}C_{ij}=\sum_{i=1}^n\sum_{j=1}^n\tilde{A}_{ij}^{(\delta_b)}\tilde{A}_{ij}^{(\delta_c)}\,;
\end{equation*}	
and the third term is 
\begin{equation*}
	\sum_{i=1}^n\sum_{j=1}^nA_{ij}C_{ij}=\sum_{i=1}^n\sum_{j=1}^nA_{ij}\tilde{A}_{ij}^{(\delta_c)}=E-\delta_c\,.
\end{equation*}	

We need to separately consider the cases in which a pair of matrices experience the same perturbation or different perturbations. In the usual case, $\delta_b \neq \delta_c$, so there is at least an offset of $|\delta_b-\delta_c|$. Consequently,
\begin{equation}\label{right}
	\sum_{i=1}^n\sum_{j=1}^n\tilde{A}_{ij}^{(\delta_b)}\tilde{A}_{ij}^{(\delta_c)}\leq E - |\delta_b-\delta_c|\,.
\end{equation}	
If, however, $\delta_b=\delta_c$ (i.e., both $\delta$-perturbations are the same), the right-hand-side of Eq.~\eqref{right} is instead given by $E$.

Altogether, this yields the following:
\begin{equation*}
	{\cal X}\leq E-\delta_b + E - |\delta_b-\delta_c| - E+\delta_c=E+(\delta_c-\delta_b)- |\delta_b-\delta_c|\,.
\end{equation*}	
Three possibilities emerge:
\begin{enumerate}
\item{If $\delta_b=\delta_c$, then ${\cal X}\leq E$.}
\item{If $\delta_b<\delta_c$, then $|\delta_b-\delta_c|= \delta_c-\delta_b$, so ${\cal X}\leq E$.}
\item{If $\delta_b>\delta_c$, then $|\delta_b-\delta_c|= \delta_b-\delta_c$, so ${\cal X}\leq E + 2(\delta_c-\delta_b)<E$.}
\end{enumerate}
This concludes the proof.
\end{proof}


\section{Appendix: Network Component Sizes} \label{appendix:components}

As we discussed in Section~\ref{scale}, half of our networks --- 30 out of 60 --- consist of two or more components after thresholding. However, even in these cases, the largest connected component of each network consists of almost the entire network. In Table~\ref{table:Components}, we show the number of components and component sizes for each of the 60 networks.

\begin{longtable}{lclc|c|}
\caption{Number of components and component sizes of each of the 60 networks. We denote treatment under Aripiprazole by ``A'', treatment under Sulpiride by ``S'', and treatment under Placebo by ``P''.
}
\label{table:Components}
\endfirsthead
\endhead
{\it Subject}    & \multicolumn{1}{l}{{\it Number of Components}} & {\it Component Sizes}                                  \\ \hline
Control 1 (A)  & 6                                        & {\{}293,1,1,1,1,1{\}}                             \\
Control 3 (A)  & 4                                        & {\{}295,1,1,1{\}}                                 \\
Control 4 (A)  & 5                                        & {\{}293,2,1,1,1{\}}                               \\
Control 5 (A)  & 20                                       & {\{}268,6,5,2,2,1,1,1,1,1,1,1,1,1,1,1,1,1,1,1{\}} \\
Control 6 (A)  & 2                                        & {\{}297,1{\}}                                     \\
Control 7 (A)  & 3                                        & {\{}296,1,1{\}}                                   \\
Control 9 (A)  & 4                                        & {\{}295,1,1,1{\}}                                 \\
Control 11 (A) & 2                                        & {\{}297,1{\}}                                     \\
Control 12 (A) & 6                                        & {\{}292,2,1,1,1,1{\}}                             \\
Control 13 (A) & 6                                        & {\{}292,2,1,1,1,1{\}}                             \\
Control 15 (A) & 2                                        & {\{}297,1{\}}                                     \\
Control 1 (P)  & 1                                        & 298                                             \\
Control 3 (P)  & 2                                        & {\{}297,1{\}}                                     \\
Control 4 (P)  & 1                                        & 298                                             \\
Control 5 (P)  & 2                                        & {\{}297,1{\}}                                     \\
Control 6 (P)  & 5                                        & {\{}289,3,3,2,1{\}}                               \\
Control 7 (P)  & 4                                        & {\{}295,1,1,1{\}}                                 \\
Control 9 (P)  & 2                                        & {\{}297,1{\}}                                     \\
Control 11 (P) & 1                                        & 298                                             \\
Control 12 (P) & 4                                        & {\{}294,2,1,1{\}}                                 \\
Control 13 (P) & 4                                        & {\{}295,1,1,1{\}}                                 \\
Control 15 (P) & 1                                        & 298                                             \\
Control 1 (S)  & 1                                        & 298                                             \\
Control 3 (S)  & 2                                        & {\{}297,1{\}}                                     \\
Control 4 (S)  & 1                                        & 298                                             \\
Control 5 (S)  & 1                                        & 298                                             \\
Control 6 (S)  & 3                                        & {\{}296,1,1{\}}                                   \\
Control 7 (S)  & 1                                        & 298                                             \\
Control 9 (S)  & 1                                        & 298                                             \\
Control 11 (S) & 2                                        & {\{}297,1{\}}                                     \\
Control 12 (S) & 1                                        & 298                                             \\
Control 13 (S) & 3                                        & {\{}296,1,1{\}}                                   \\
Control 15 (S) & 1                                        & 298                                             \\
Patient 1 (A)  & 2                                        & {\{}297,1{\}}                                     \\
Patient 2 (A)  & 1                                        & 298                                             \\
Patient 4 (A)  & 1                                        & 298                                             \\
Patient 6 (A)  & 1                                        & 298                                             \\
Patient 7 (A)  & 2                                        & {\{}297,1{\}}                                     \\
Patient 8 (A)  & 7                                        & {\{}291,2,1,1,1,1,1{\}}                           \\
Patient 9 (A)  & 1                                        & 298                                             \\
Patient 10 (A) & 1                                        & 298                                             \\
Patient 12 (A) & 1                                        & 298                                             \\
Patient 1 (P)  & 1                                        & 298                                             \\
Patient 2 (P)  & 1                                        & 298                                             \\
Patient 4 (P)  & 1                                        & 298                                             \\
Patient 6 (P)  & 1                                        & 298                                             \\
Patient 7 (P)  & 2                                        & {\{}297,1{\}}                                     \\
Patient 8 (P)  & 7                                        & {\{}292,1,1,1,1,1,1{\}}                           \\
Patient 9 (P)  & 1                                        & 298                                             \\
Patient 10 (P) & 3                                        & {\{}296,1,1{\}}                                   \\
Patient 12 (P) & 1                                        & 298                                             \\
Patient 1 (S)  & 1                                        & 298                                             \\
Patient 2 (S)  & 2                                        & {\{}297,1{\}}                                     \\
Patient 4 (S)  & 1                                        & 298                                             \\
Patient 6 (S)  & 1                                        & 298                                             \\
Patient 7 (S)  & 1                                        & 298                                             \\
Patient 8 (S)  & 5                                        & {\{}294,1,1,1,1{\}}                               \\
Patient 9 (S)  & 1                                        & 298                                             \\
Patient 10 (S) & 1                                        & 298                                             \\
Patient 12 (S) & 1                                        & 298                                            \\
\hline
\end{longtable}

\end{appendix}

 
 


\end{document}